\documentclass[amsmath,amssymb,showpacs,floatfix,reprint,jcp,numerical]{revtex4-1}
\usepackage{graphicx}
\usepackage{bm}
\usepackage{hyperref}
\usepackage{color}
\begin{document}

\title{Glassy dynamics of dense particle assemblies on a spherical substrate}
\author{Julien-Piera Vest,  Gilles Tarjus, and Pascal Viot}
\affiliation{$^1$Laboratoire de Physique Th\'eorique de la Mati\`ere Condens\'ee, CNRS  UMR 7600, UPMC-Sorbonne Universit{\'e}s,
 4, place Jussieu, 75252 Paris Cedex 05, France}

\begin{abstract}
We study by Molecular Dynamics simulation a dense one-component system of particles confined on a spherical substrate. We more specifically investigate the evolution of the structural and dynamical properties of the system when changing the control parameters, the temperature and the curvature of the substrate. We find that the dynamics becomes glassy at low temperature, with a strong slowdown of the relaxation and the emergence of dynamical heterogeneity. The prevalent local $6$-fold order is frustrated by curvature and we analyze in detail the role of the topological defects in the statics and the dynamics of the particle assembly. 
\end{abstract}
\date{\today}
\newcommand{\thedate}{\today}
\maketitle

\section{Introduction}

In a variety of physical situations dense assemblies of particles form on curved spherical substrates. Particles are irreversibly adsorbed to the substrate but remain mobile at the surface so that they can in principle reach equilibrium phases. This takes place in adsorption and coating phenomena  \cite{Post1986,Post1988} or in particle-stabilized emulsions \cite{Aveyard2003,Tarimala2004,Bausch2003,Einert2005,Lipowsky2005,Irvine2010,Leunissen2007,Irvine2012,Folter2012,Qi2014,Sicard2016}. In the latter case, wetting and/or electrostatics at the interface between two immiscible liquids lead to an essentially irreversible adsorption of colloidal particles at the interface. As a result, the two liquids, {\it e.g.}, oil and water, can form a stable emulsion, as known from the early work of Pickering \cite{Pickering1907}. The colloidal particles coat the drops of one liquid immersed in the other one in the form of a layer, and it is the presence of these layers that stabilizes the emulsion.

The properties of these particle assemblies depend on density and temperature. They also depend on the curvature of the substrate which, in the present case, is constant and positive. Most of the existing studies have focused on the high-density and low-temperature regime where the phase of the adsorbate appears essentially crystalline and, correspondingly, the mobility of the particles is very restricted on the observation time. These ``spherical crystals'' are hexagonal structures interrupted by topological defects forming finite-length strings known as ``grain boundary scars''  \cite{Bausch2003,Einert2005,Lipowsky2005,Perez-Garrido1997,Bowick2000,Bowick2007,Sausset2010,Irvine2010,Guerra2018a}. Here, we rather address what is the structural and the dynamical behavior of the particle layer when density is lowered and/or temperature is raised while maintaining a condensed, {\it i.e.}, liquid-like, phase.

At a more abstract level, spherical substrates have been used as a trick in computer simulations of systems involving long-range interactions and/or long-range spatial correlations in standard Euclidean space. In such a case, the numerical investigation requires appropriate boundary conditions and the thermodynamic limit may be difficult to approach. ``Spherical boundary conditions'', which amount to considering the system on the surface of a sphere (or a hypersphere) where no boundaries need to be specified and increasing the radius of this sphere, have then been proposed as an alternative to the common procedure \cite{Perez-Garrido1998,Moore1999,Hansen1979,Caillol1991,Caillol1992,Caillol1999,Prestipino2012}

On a general theoretical ground, studying the properties of a system in uniformly curved space provides an extra control parameter (the Gaussian curvature) in addition to the common thermodynamic parameters. This may prove interesting in several cases. For instance, curving space may introduce  some ``geometric frustration''. This is generally the case for two-dimensional assemblies of spherical particles. In flat (Euclidean) space and for high enough density and low enough temperature the local arrangement of the particles has a prevalent $6$-fold character that leads to hexatic and hexagonal phases at large distance \cite{Nelson1983}. However, a constant nonzero curvature thwarts the long-range or quasi-long-range ordering by introducing an irreducible density of topological defects in the particle configurations \cite{Bowick2000,Nelson2002,Tarjus2005,Bowick2007,Sausset2008c,Sausset2010,Sausset2010a,Tarjus2012}. The $6$-fold hexatic and hexagonal orders are  therefore frustrated and the degree of frustration is controlled by the curvature of space. According to the frustration-based theory of glass-forming liquids \cite{Nelson2002,Tarjus2005} this can lead to glass formation and, correspondingly, to a slow, glassy, dynamics as density increases or temperature decreases. This idea has been tested in detail in the case of a two-dimensional space of constant negative Gaussian curvature, {\it i.e.} the hyperbolic plane $H^2$ \cite{Sausset2008c,Sausset2010,Sausset2010a}. It is then worth investigating here if a similar behavior is found in the case of a two-dimensional manifold of constant positive Gaussian curvature, {\it i.e.} the sphere $S^2$.

In this paper, we study by Molecular Dynamics simulation a one-component system of spherical particles confined on the sphere $S^2$.  The geometry is considered as frozen, providing a background whose metric and topological characteristics affect the behavior of the embedded system, but with no feedback from the latter. We more specifically investigate the evolution of the structural and dynamical properties of the system when changing the control parameters. We consider the variation of temperature at a constant (high) density and the variation of the curvature of the substrate (which for a compact manifold as the sphere is related to the system size). 

Our main focus is the dynamics and the relaxation to equilibrium of the particle assembly, a feature that has not received much attention thus far
(see however [\onlinecite{Lipowsky2005,Bowick2007}]). In addition to the connection to the problem of the glass transition and its frustration-based theoretical approach, establishing the behavior of the typical equilibration time, the associated effective activation energy barriers at low temperature, and the main dynamical mechanisms is of relevance if one wishes to vary the range of temperature and density over which particle assemblies are used in practical applications \cite{Ling2005}.

The outline of the paper is as follows. In Section 2, we introduce the model and some aspects of the Molecular Dynamics simulation on the sphere. We also introduce the structural and dynamical quantities that we have monitored. In Section 3, we present the predictions of the continuum elasticity theory of topological defects on the sphere, first developed by Nelson, Bowick, Travesset, and coworkers \cite{Bowick2000,Bowick2007,Bowick2009}. We show and   discuss the results for the structural quantities in Section 3 and for the dynamics in Section 4. We finally provide some concluding remarks in Section 5.

\section{Dense liquid on a sphere: Model and method}

\subsection{Model and simulation}

We consider a monodisperse system  of $N$ spherical particles living on the surface of a $3$-dimensional sphere of radius $R$ (a $2$-sphere $S^2$) and interacting through a pairwise Lennard-Jones potential
 \begin{equation}
 v(r)=4\epsilon\left[\left(\frac{\sigma}{r}\right)^{12}-\left(\frac{\sigma}{r}\right)^6\right]\,, 
 \end{equation}
 where $r$ denotes  the geodesic distance between two particle centers. The Lennard-Jones potential is truncated after a cutoff distance $r_c=2.5\sigma$, which guarantees for the values of $R$ considered in this study that a particle does artificially interacts with itself (due to the compactness of $S^2$); $m$, $\sigma$, and $\epsilon$ are the particle mass, diameter, and the energy scale, while the unit of time is taken as $\sqrt{m\sigma^2/\epsilon}$. The curvature of the substrate introduces another intrinsic length scale $R$ in the problem and the dimensionless ratio $R/\sigma$ is now an additional control parameter. On the surface of a sphere of radius $R$, the reduced density is fixed by the number of particles and the ratio $R/\sigma$ according to
\begin{equation}
\label{eq_reduced_density}
\tilde \rho=\frac{2N}{\pi}\left[1-\cos\left(\frac{\sigma}{2R}\right)\right]\,,
\end{equation}
which in the Euclidean (flat) limit, $R\rightarrow \infty$, goes to the usual definition of the reduced density, $\tilde \rho\rightarrow (N/A)\sigma^2$, 
where $A$ is the  area of the system.

We have chosen the Lennard-Jones potential to allow a direct comparison with the previous study of a dense liquid on the hyperbolic plane, a surface of constant negative Gaussian curvature \cite{Sausset2008c,Sausset2010,Tarjus2012} and investigate the role of the sign of the curvature (and the compactness of space) on the system's properties.  This potential is {\it a priori} not the best suited for describing the interactions between colloidal particles at fluid interfaces. For repulsive colloids, a hard-sphere potential, possibly supplemented by a Yukawa interaction accounting for screened electrostatic interaction such as the double-layer repulsion form of the DLVO theory \cite{russel1989} or for capillary interactions mediated by the liquid interface \cite{Kralchevsky2000,Ershov2013,Liu2018}, would be more appropriate. (Charged hydrophobic colloidal particles adsorbed at oil-aqueous interfaces and interacting through a dipolar-like pair potential have also been considered, see Ref. [\onlinecite{Guerra2018a}].) However, it was shown \cite{Kusumaatmaja2013} that the detailed form of the potential has no qualitative effect on the configurations observed on curved substrates provided the particle density is high enough, which is the case here. We therefore expect that the Lennard-Jones liquid will qualitatively share the structural and dynamical properties of colloidal assemblies on spherical substrates. This may not be true for the case of attractive colloids with a hard core plus a very short-ranged attractive interaction due to depletion forces \cite{Meng2014}, whose behavior is significantly different from other colloidal systems and is dominated by dendritic-like structures, but we do not consider such kind of frustration-induced self-limited growth \cite{Grason2016} in this work.

We study the structural and dynamical properties of the system by a Molecular Dynamics simulation. To do this it is convenient to view each particle as a $3$-dimensional rotator rigidly linked to the center of the sphere so that it is constrained to move at a fixed distance $R$ of this center. The Hamiltonian is then 
\begin{equation}
 H=\sum_{i=1}^N \frac{m{\boldsymbol \omega}_i^2}{2mR^2}-\frac{1}{2}\sum_{i\neq j}{v(r_{ij})}\,,
\end{equation}
where ${\boldsymbol \omega}_i$ is the angular velocity of the rotator $i$, and the associated equations of motion read 
\begin{equation}
 m\dot{{\boldsymbol \omega}}_i=-{\bf r}_i\times \sum_{j\neq i}\partial_ {\bf r_i} v(r_{ij})\,,
\end{equation}
with $r_{ij}$ the geodesic distance between rotators $i$ and $j$. A Molecular Dynamics simulation can then be implemented via a ``velocity Verlet algorithm''. Details concerning the implementation and the preparation of the initial configurations are given in Refs. [\onlinecite{vest2014,vest2015}]. 

\subsection{Measured structural and dynamical quantities}

To describe the structure of the dense liquid on $S^2$, we have computed the pair correlation function $g(r)$ associated with the probability of finding another particle center at a geodesic distance $r$ of a given one as well as the (static) bond-orientational pair correlation function $G_6(r)$ that characterizes the extension of the $6$-fold local order. The latter is obtained as follows. For a particle $j$, one can define the local bond-orientational order parameter, $\Psi_6(j)=(1/N_{b,i})\sum_{k/j} \exp(6i\theta_{jk})$, where  the sum over $k$ runs over the nearest neighbors of the particle $j$, $N_{b,j}$ is the number of such nearest neighbors, and $\theta_{jk}$ is the angle characterizing the ``bond''  between $j$ and $k$. The nearest neighbors and the ``bonds'' are defined through a Voronoi-Delaunay tesselation of the particle configurations on the $2$-sphere (see below). In order to measure the extension of this local order, we consider the pair correlation $G_6(r)$,
\begin{equation}
 G_6(r)=\frac{1}{Ng(r)}\sum_{j,k}\langle \Psi_6^*(j)\Psi_6(k)\rangle \frac{\delta(r_{jk}-r)}{2\pi  R\sin(r/R)}
\end{equation}
where  the brackets denote the average over equilibrium configurations and $r_{jk}$ is the geodesic distance between particles $j$ and $k$.

The dynamics of  the system has been analyzed by using the time-dependent $2$-point correlation functions of the density fluctuations: the self-intermediate scattering function $F_s(k,t)$ and the full intermediate scattering function $F(k,t)$.
On $S^2$ the former is given by \cite{Tarjus2012}
\begin{align}
F_s(k,t)&=\frac{1}{N}\sum_{j} \bigg<P_k\left [\cos\left(\frac{d_{j}(0,t)}{R}\right )\right ]\bigg>\,,
\label{eq_Fs}
\end{align}
where $d_{j}(0,t)$ is the geodesic distance between the positions of particle $j$ at times $0$ and $t$,  $k$ is an integer, and  $P_{k}$ is the $k$th Legendre polynomial. The latter is given by 
\begin{align}
F(k,t)&=\frac{1}{N}\sum_{i, j} \bigg<P_k\left [\cos\bigg(\frac{d_{jk}(0,t)}{R}\bigg)\right ]\bigg>\,,
\label{eq_Fcoll}
\end{align}
where  $d_{jk}(0,t)$ is the geodesic distance between the position of particle $j$ at time $0$ and that of particle $k$ at time $t$. In order to compare the time dependences of the two intermediate scattering functions,  $F(k,t)$ can be normalized by the static structure factor $S(k)$, which is the Fourier transform on the sphere of $g(r)-1$ \cite{Sausset2009,vest2015}: then, $F(k,t=0)/S(k)=1=F_s(k,t=0)$.

In dense two-dimensional particle assemblies it is also useful to consider a dual description in terms of configurations of topological defects in the underlying $6$-fold hexatic or hexagonal order. These point-like defects can be defined at a microscopic level through a Voronoi tesselation or its dual, the Delaunay triangulation. These constructions allow one to uniquely determine the number of nearest neighbors of any given particle. A ``disclination" corresponds to a particle having strictly less (positive disclination) or more (negative disclination) than 6 nearest neighbors and is a defect in the $6$-fold (hexatic) bond-orientational order. A ``dislocation" on the other hand appears as a ``dipole'' formed by a positive and a negative disclination and breaks translational order. We have implemented the Delaunay triangulation of the configurations on $S^2$ with the Stripack library \cite{Renka1997} [the computation scales as O($N\ln N$)].

In curved space, topological defects appear not only as thermal excitations but also as an intrinsic consequence of the geometric frustration, {\it i.e.}, the mismatch between (quasi) long-range $6$-fold order and curvature. The topology of the embedding manifold indeed constrains the number of defects. In two dimensions, this straightforwardly derives from the Euler-Poincare relation,
\begin{equation}
\frac{N}{6}(6-\overline{z})=\chi,
\label{eq_euler-poincare}
\end{equation}
where $\overline z$ is the mean coordination number of the particles and $\chi$ is the Euler characteristic of the manifold, which is equal to 0 for the Euclidean plane, 2 for the sphere $S^2$, and $\leq -2$ in a hyperbolic geometry. As a result, on $S^2$ there must be an excess of particles with less than 6 neighbors, \textit{i.e.} of positive disclinations. The minimum number is 12 disclinations of ``topological charge" $+1$ (12 particles with 5 neighbors) in an otherwise 6-fold coordinated medium, which then fulfills the constraint of Eq. (\ref{eq_euler-poincare}). There may be more defects provided the sum of their topological charge is zero, {\it e.g.}, a distribution of dislocations in addition to the $12$ ``irreducible'' $+1$ disclinations. 

Although experiments on colloids are more conveniently performed by controlling the density $\rho$ at fixed temperature $T$ (see also the experimental study in Ref. [\onlinecite{Guerra2018a}], where the control parameter is a dimensionless quantity, $\Gamma \propto \rho^{3/2}/T$), we have chosen to vary the temperature and fix the density at a high reduced value $\tilde \rho=0.92$ typical of a liquid phase. As mentioned before, this allows us to directly compare our results to those previously obtained on the hyperbolic plane. Furthermore, varying the temperature gives us a more direct access to the effective activation energy barriers controlling the dynamics. The qualitative trends when decreasing the temperature or increasing the density should moreover be similar. We have also changed the curvature of the substrate and studied a range of dimensionless radius of curvature: $R/\sigma \simeq 7.2$, $9.3$, $13.2$, $18.6$, $26.3$, and $32.2$. For the chosen reduced density this corresponds to system sizes $N=600$, $1000$, $2000$, $4000$, $8000$, and $12000$, respectively: This is summarized in Table \ref{tab:table1}. In addition we have carried out simulations in the Euclidean plane (with rectangular periodic boundary conditions) for reference.

\begin{table}[b]
\caption{\label{tab:table1}%
Correspondence between the total number of particles $N$ and the dimensionless radius of curvature $R/\sigma$ for a reduced density $\tilde \rho=0.92$ on the $2$-sphere $S^2$.
}
\begin{ruledtabular}
\begin{tabular}{lc}
\textrm{$N$}&
\textrm{$R/\sigma$}
\\
\colrule
$600$&$7.2$  \\
 $1000$ &$9.3$ \\
 $4000$& $18.6$ \\
$8000$&$26.3$  \\
 $12000$ &$32.2$   
\end{tabular}
\end{ruledtabular}
\end{table}

\section{Continuum elasticity theory of defects on the sphere}

\subsection{Continuum theory of defects} 
  
We briefly summarize below the continuum elastic theory of defects in a hexagonal medium on the sphere $S^2$ developed by Nelson, Bowick, Travesset and coworkers \cite{Bowick2000,Bowick2007,Bowick2009}. The main ingredient is a free energy functional formulated for an arbitrary disclination density in an underlying hexagonal medium on a $2$-dimensional curved manifold, which is valid at large length scale and low temperature:
\begin{align}
F&=\frac{Y}{2} \iint d^2{\bf r}d^2{\bf r}'\sqrt{g({\bf r})}\sqrt{g({\bf r}')}\big[K({\bf r})-Q({\bf r})\big]\frac{1}{\Delta^2}\bigg|_{{\bf r},{\bf r}'}\nonumber\\
&\times \big[K({\bf r}')-Q({\bf r}')\big] +N_D E_{D,core},
\label{energie_defauts}
\end{align}
where $g({\bf r}$ is the  determinant of the metric tensor (on $S^2$, $g({\bf r}=(\theta,\phi))=[R^2\sin(\theta)]^2$), $K({\bf r})$ is the Gaussian curvature at point ${\bf r}$ (on $S^2$, $K=1/R^2$), $Y$ is the Young modulus of the hexagonal crystal, $1/\Delta^2$ is the (operator) inverse of the squared Laplacian, $N_D$ the total number of disclinations, and $E_{D,core}$ is the mean renormalized core energy of the disclinations; finally, $Q({\bf r})$ is the disclination density defined as
\begin{equation}
Q({\bf r})=\frac{\pi}{3\sqrt{g({\bf r})}} \sum_{i=1}^{N_D}q_i \delta({\bf r}-{\bf r}_i),
\label{eq:59}
\end{equation} 
where $q_i$ is the topological charge of the $i$th disclination.

\begin{figure}[t]
\begin{center}
\resizebox{8cm}{!}
 {\includegraphics{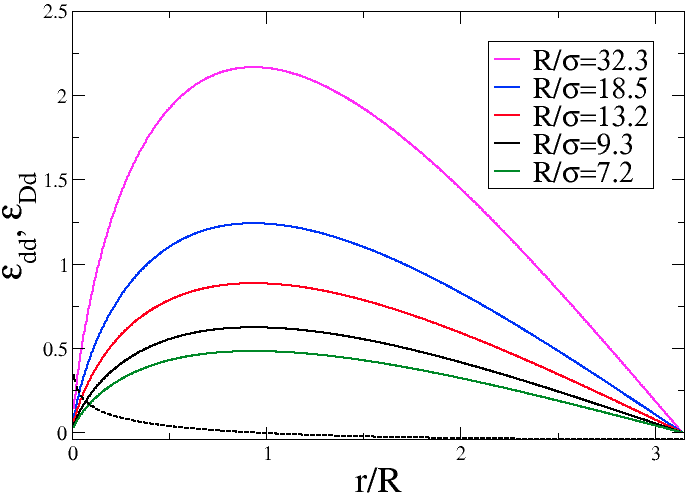}}
 \caption{Interaction energies between topological defects in a geodesically aligned grain boundary scar, $\epsilon_{Dd}$ and $\epsilon_{dd}$, as a function of the dimensionless geodesic distance  $r/R$. The dislocation-dislocation pair interaction energy $\epsilon_{dd}$ is a monotonously decreasing function of  $r/R$ (dashed line). The disclination-dislocation pair interaction energy, $\epsilon_{Dd}$, on the other hand has a nonmonotonous behavior and goes through a maximum for $r\approx 0.94 R$; it is plotted for  different curvatures: from bottom to top, $R/\sigma=7.2$, $9.3$, $13.2$, $18.6$, and $32.2$.}
\label{fig_energie_defauts}
\end{center}
\end{figure}

After dividing the disclinations into the minimum number required by the topological constraint ($12$ on the sphere) and the others that form dislocation dipoles (see above) and performing some additional manipulations \cite{Bowick2000}, one obtains the free energy in the form of  two contributions: one corresponds to the (pairwise) interaction between disclinations and dislocations, $F_{Dd}$, and the other to the (pairwise) interaction between dislocations, $F_{dd}$. The expressions are given in Ref. [\onlinecite{Bowick2000}] and are not reproduced here. It is nonetheless instructive to plot the corresponding pair interaction terms $\epsilon_{Dd}$ and $\epsilon_{dd}$ in the simple geometry of a geodesically straight ``boundary scar'' formed by one $+1$ disclination and a string of dislocations: see Fig.~\ref{fig_energie_defauts}. The interaction between dislocations is always repulsive, whereas the interaction between a disclination and a dislocation is  attractive for $r\lesssim R$ and repulsive for $r\gtrsim R$. This is the origin of the presence of grain boundary scars even in the ground state of the system. The theory leads to a prediction concerning the number of dislocations per irreducible disclinations at low temperature
\begin{equation}\label{eq_GBSlength}
N_d/12 \approx  0.41 (R/\sigma) - cst \,.
\end{equation}
This prediction has already been successfully checked in experimental \cite{Bausch2003}  and numerical \cite{Backofen2010} studies.

\subsection{Relaxation due to dislocation dynamics}

At low temperatures one may expect that the relaxation of the system is controlled by the dynamics of the topological defects. Since the structure of the system on $S^2$ at such temperatures takes the form of a hexagonal medium intertwined by $12$ grain boundary scars separated roughly by a distance of the order of $R$, the dominant contribution to the long-time dynamics should be given by dislocation glide from one scar to another. 

Dislocation glide for times shorter than the relaxation time of the system has been theoretically and experimentally studied in Refs. [\onlinecite{Bowick2007,Lipowsky2005}]. On the other hand, a scenario for the long-time dynamics of equilibrium dense particles assemblies in the negatively curved hyperbolic plane has been proposed in Ref. [\onlinecite{Sausset2010a}]. Dislocation glide is an activated process at low temperature and, according to this description, the activation barrier includes two contributions: one is due to the Peierls potential associated with the underlying hexagonal medium and one comes from the energy cost due to a weakening of the disclination-dislocation interaction when a dislocation glides away from a given boundary scar and migrates over a distance of the order of the radius of curvature, here $R$. As a result one predicts the following dependence of the relaxation time on the curvature:
\begin{equation}
\ln(\tau/\tau_\infty)\sim \frac{E_0}{kT}+\frac{CY\sigma R}{kT}\sim \frac{E_0}{kT}+\frac{E_1(R/\sigma)}{kT},
\label{eq_tau}
\end{equation}
where $E_0$ is the barrier corresponding to Peierls potential and  $E_1=C Y\sigma^2$ with $C$ a constant and $Y$ the Young modulus.  For large enough system size, $R/\sigma \gg 1$, the effective barrier to relaxation, $\Delta E$, should then have a linear behavior  on $R/\sigma$ at low temperature, which means a square-root dependence on the total number of particles,
\begin{equation}
\Delta E \propto R/\sigma\sim \sqrt{N}\,.
\label{eq_delta_E_dislo}
\end{equation}
In Ref. [\onlinecite{Sausset2010a}] a scaling argument was further invoked to predict that when increasing the temperature the relaxation time should behave as in Eq. (\ref{eq_tau}) with the radius of curvature $R$ replaced (up to a multiplicative constant of order one) by the bond-orientational correlation length $\xi_6(T)$ obtained from the correlation function $G_6(r)$ introduced above.

\begin{figure}[t]
\begin{center}
 \resizebox{8.5cm}{!}{\includegraphics{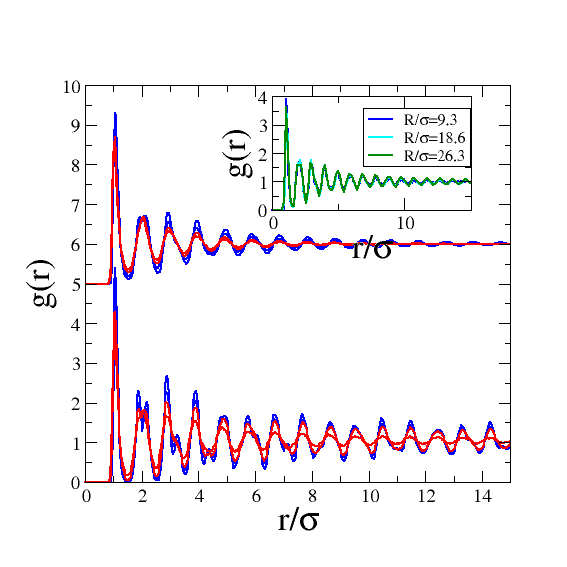}}
 \caption{Pair correlation function $g(r)$ as a function of the (geodesic) distance $r$ for a Lennard-Jones system of $N=1000$ particles at a reduced density $\tilde \rho = 0.92$ and for temperatures $T=2,1.5,1.2,0.8$. The red curves correspond to  $T>T^*$ (where $T^*\approx 1.3-1.4$ is the ordering temperature of the system in $E_2$) and  the blue curves to $T<T^*$. The top curves are for the sphere $S^2$ (result shifted by a step of $5$), for which $R/\sigma\approx 9.3$, and the bottom curves are for the Euclidean plane $E^2$ with periodic boundary conditions. The insert shows the pair correlation function at $T=1.0$ for three different curvatures.}
\label{fig_radial_function}
\end{center}
\end{figure}
\section{Results for the statics}

We now discuss the results obtained from our computer simulation and we start with those describing the structural properties of the particle assembly.

\subsection{Static correlation functions}

For completeness we first present the results for the pair correlation function $g(r)$ for various temperatures and curvatures.  We illustrate them for $R/\sigma=9.3$ ($N=1000$) in Fig.~\ref{fig_radial_function} together with the behavior of the same system on the Euclidean plane. As can be seen there is no signature of long-range nor quasi-long-range translational order on $S^2$, contrary to what is found in flat space: The correlation length is never larger than, say, $5\sigma$, and does not vary with temperature in any significant way. Moreover, even at low temperature where the system is ordered in the Euclidean plane, the change in $g(r)$ with curvature is small (see the inset of Fig.~\ref{fig_radial_function}). 
For further use, we note that the ordering temperature in flat space (irrespective of the details about the nature and the number of transitions) is at about $T^*\approx 1.3-1.4$ for the chosen reduced density $\tilde \rho=0.92$.

\begin{figure}[t!]
\begin{center}
  \resizebox{8cm}{!}
  {\includegraphics{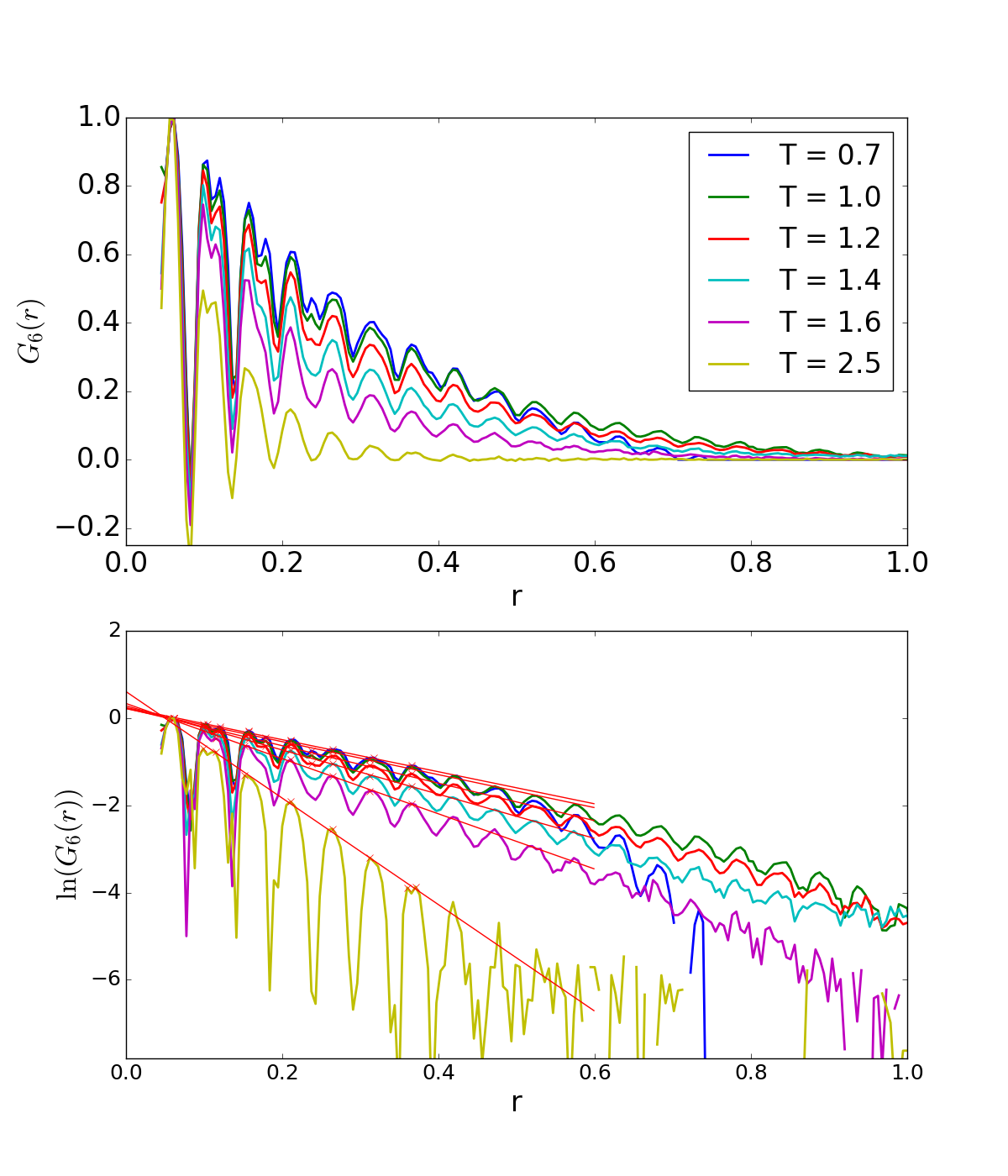}}
   \caption{\label{fig:G6} Pair correlation function $G_6(r)$ of the $6$-fold bond-orientational order parameter at different temperatures for a reduced radius of curvature $R/\sigma\approx 18.6$ ($N=4000)$.Top: linear-linear plot.  Bottom: log-linear plot where the red lines indicate an exponential behavior that allows us to extract a correlation length $\xi_6$ for each temperature.}
\end{center}
\end{figure}

More significant is the behavior of the bond-orientational correlation function associated with $6$-fold local order. In Fig.~\ref{fig:G6} we illustrate the evolution of $G_6(r)$ when changing the temperature for a radius of curvature $R/\sigma \approx 18.6$ ($N=4000$). The decay of the function versus the geodesic distance $r$  becomes significantly slower as temperature decreases, showing an increase of 6-fold bond-orientational spatial correlations. In the bottom part of Fig.~\ref{fig:G6} the correlation function is displayed in a log-linear plot shows and one can see that the envelope of the function can be reasonably well approximated by an exponential, which allows us to extract a correlation length, $\xi_6$.

\begin{figure}[t]
\begin{center}
   \resizebox{8.5cm}{!}
   {\includegraphics{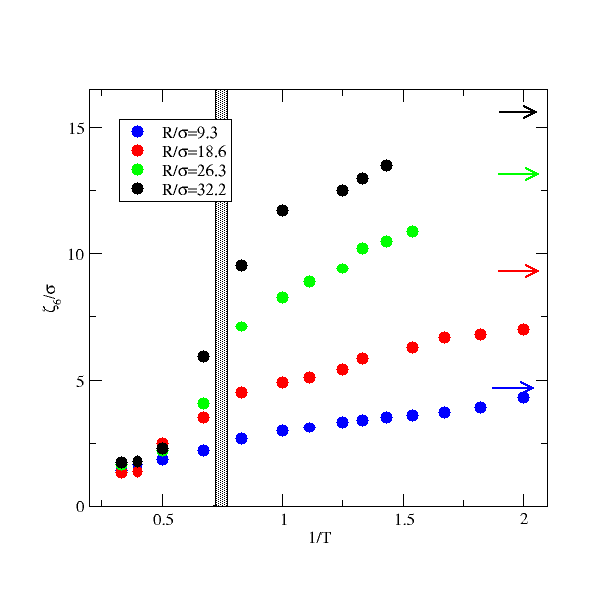}}
   \caption{\label{fig:xi6} Dimensionless bond-orientational correlation length $\xi_6/\sigma$  versus  the inverse of the temperature for different curvatures: $R/\sigma \approx 9.3$, $18.6$, $26.3$, $32.2$.   The arrows indicate the values $R/(2\sigma)$ and we have also shown the approximate location of the ordering temperature in flat space, $1/T^* \approx 0.72-0.77$ (marked as a shaded area).}
\end{center}
\end{figure}

The temperature dependence of the bond-orientational correlation length $\xi_6$ in units of $\sigma$ is displayed in Fig.~\ref{fig:xi6} for four different curvature ratios $R/\sigma$. One observes a monotonous increase of $\xi_6$ when $T$ is lowered with a tendency to saturate at the lowest temperature.  (This is more easily seen for the largest curvatures for which we can equilibrate the system at lower temperature.) For all curvatures, the high-temperature limit of $\xi_6$ is equal to $1.5$ to $2 \sigma$, and $\xi_6$ appears to saturate at low temperature at a value close to $R/2$. This is a consequence of the geometric frustration that prevents an extension of the $6$-fold order over the whole spherical substrate. The typical distance between the 12 irreducible disclinations is of the order of the radius $R$ (for disclinations regularly arranged on the vertices of an icosahedron, the distance between neighboring disclination is about $1.05 R$), which is compatible with the saturation values of $\xi_6$. One can see, especially for the smallest curvatures, that the increase of $\xi_6$ is more rapid  in a temperature interval close to the ordering temperature in the euclidean plane ($1/T^*\approx 0.72-0.77$). This illustrates the phenomenon of a frustration-induced  avoided transition \cite{Tarjus2005,Esterlis2017}. In a sense, the system on $S^2$ can be considered as a ``supercooled" liquid below $T^*$.

\subsection{Topological defects}

\begin{figure}[h!]
\begin{center}

\includegraphics[width=0.15\textwidth]{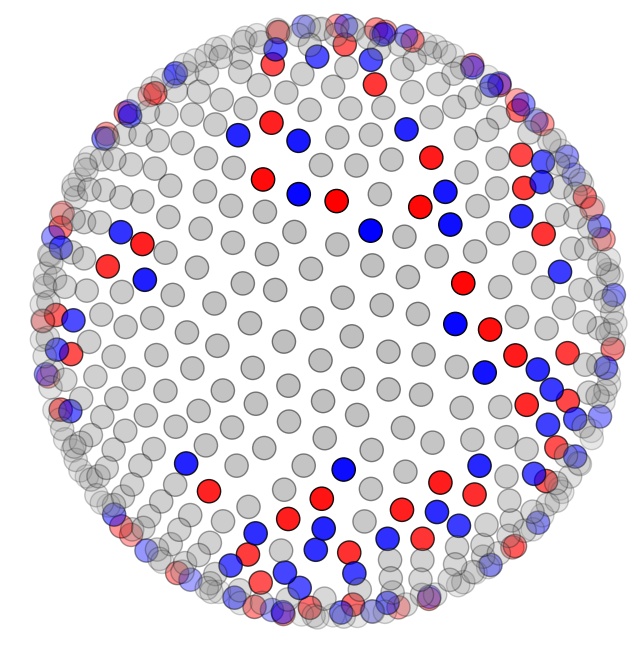}
  \includegraphics[width=0.15\textwidth]{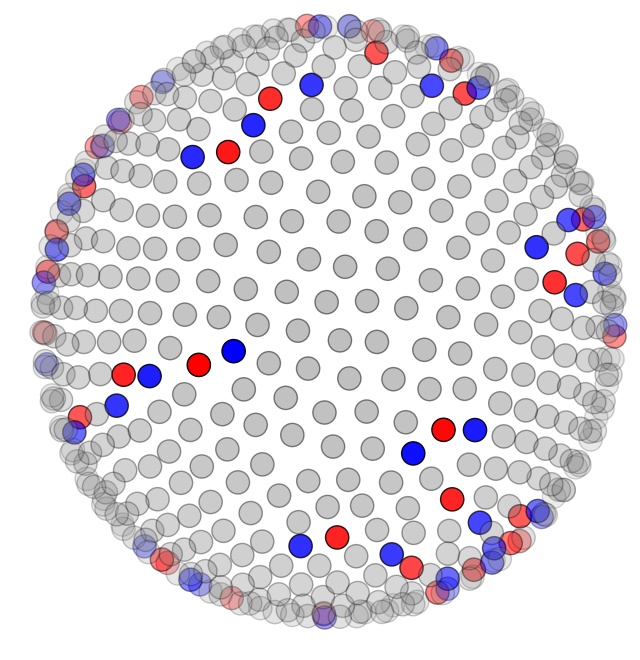}
  \includegraphics[width=0.15\textwidth]{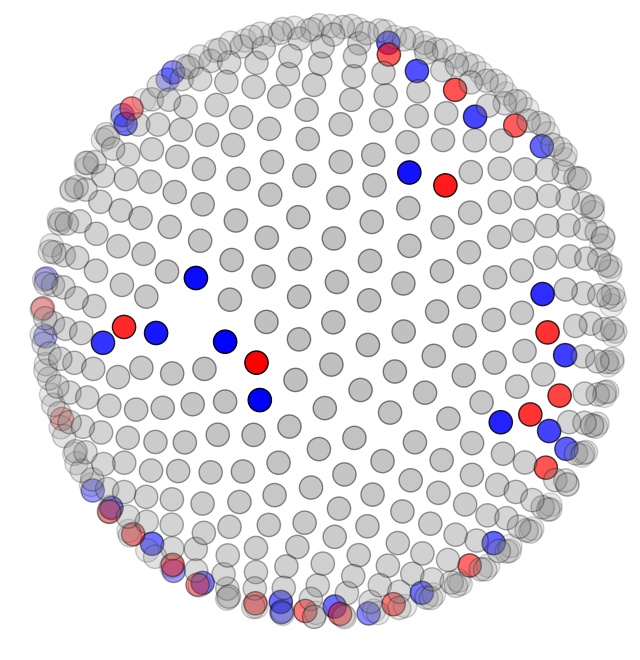} 
  \vspace{2mm}
  \includegraphics[width=0.15\textwidth]{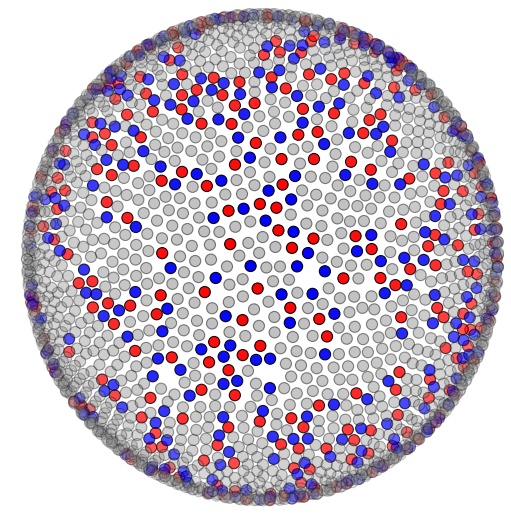}
  \includegraphics[width=0.15\textwidth]{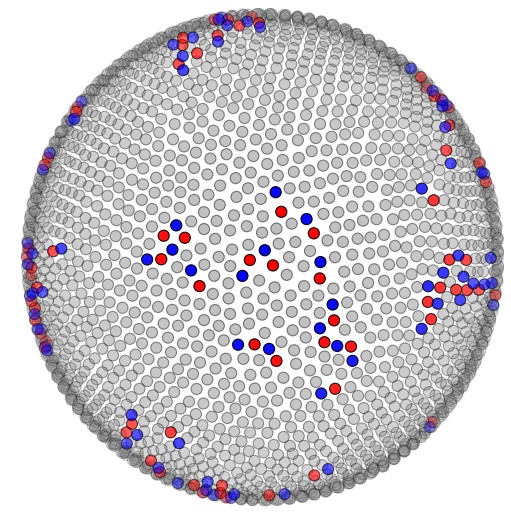}
  \includegraphics[width=0.15\textwidth]{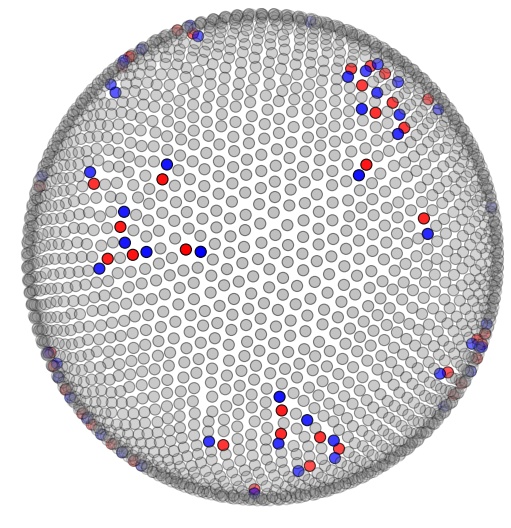} 
   \vspace{2mm}
    \includegraphics[width=0.15\textwidth]{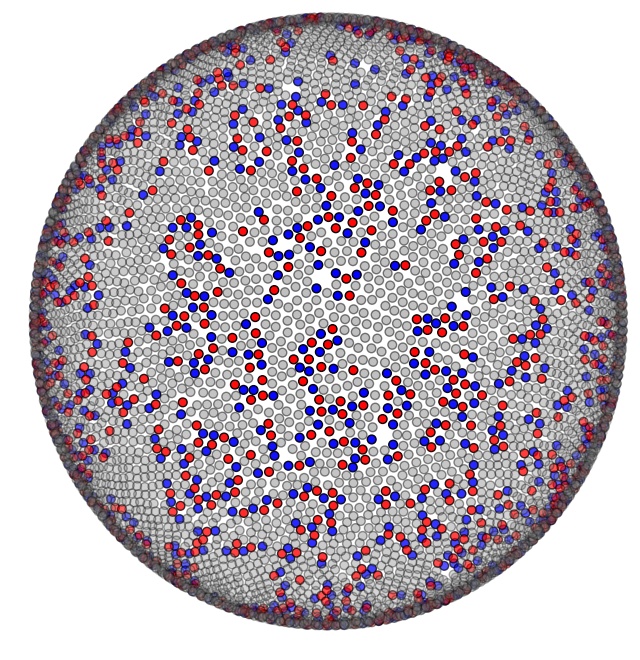}
    \includegraphics[width=0.15\textwidth]{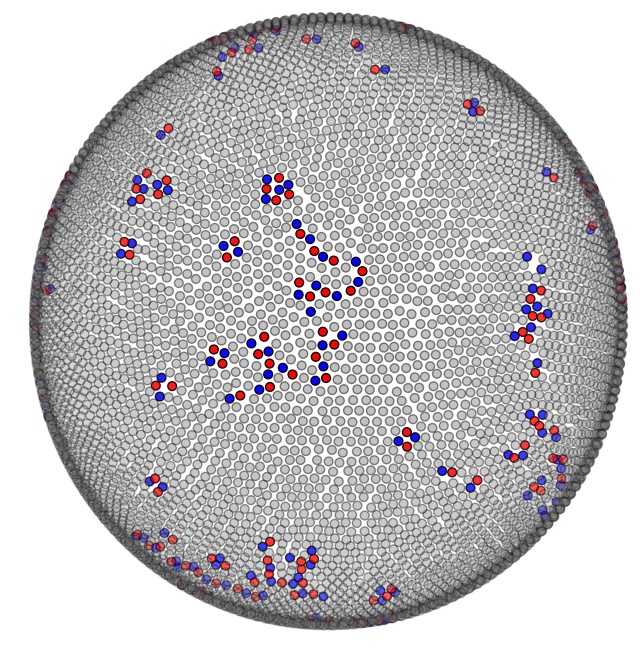}
    \includegraphics[width=0.15\textwidth]{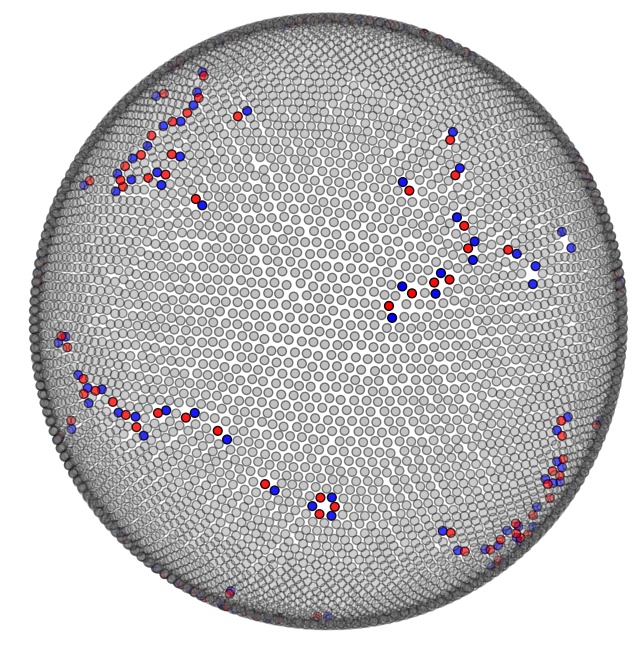}
   \caption{ \label{fig:config_defauts} Typical equilibrium particle configurations for a density $\tilde{\rho}=0.92$, reduced values of the radius of curvature  $R/\sigma=9.3$ (top), $13.2$ (middle), and $32.2$ (bottom), for three temperatures  $T=2$, $1$, and $0.7$ (from left to right). Gray particles have $6$ neighbors, red particles have $7$ neighbors and correspond to negative $-1$ disclinations, while blue particles have $5$ neighbors and correspond to positive $+1$ disclinations. There are very few particles with more than $7$ (red) or less than $5$ (blue) neighbors .}
\end{center}
\end{figure}

Typical equilibrium particle configurations are displayed in Fig.~\ref{fig:config_defauts} at the reduced density $\tilde{\rho}=0.92$ for three temperatures,  $T=2$ (liquid state), $1$ (weakly  ``supercooled" liquid), and $0.7$ (strongly ``supercooled" liquid), and for three values of the reduced radius of curvature $R/\sigma=9.3$, $13.2$, and $32.2$. The color code is as follows: $6$-fold coordinated particles ($6$ neighbors) are in grey,  $5$- and $7$-fold coordinated particles are in blue and red, respectively, and correspond to positive ($+1$) and negative ($-1$) disclinations. At high temperature, defects are numerous, due to thermal fluctuations, and they are more or less randomly distributed on the sphere. Below the avoided transition, at $T=1$, the number of defects has significantly dropped, and one can already clearly see that they tend to organize in localized structures, leaving the rest of the particle assembly with a $6$-fold local order. This is the dual picture of the increase of the bond-orientational correlation length shown in Fig.~\ref{fig:xi6}.

The disclinations of opposite signs (red and blue particles) tend to pair to form topologically neutral dislocations. At the lowest temperature $T=0.7$, the configuration of the defects  evolves towards $12$ chains (not necessarily geodesically aligned) comprising a number of dislocations and one excess $+1$ disclination. The typical chain size appears to increase with  $R/\sigma$. 

\begin{figure}[ht]
\begin{center}
    \includegraphics[width=0.48\textwidth]{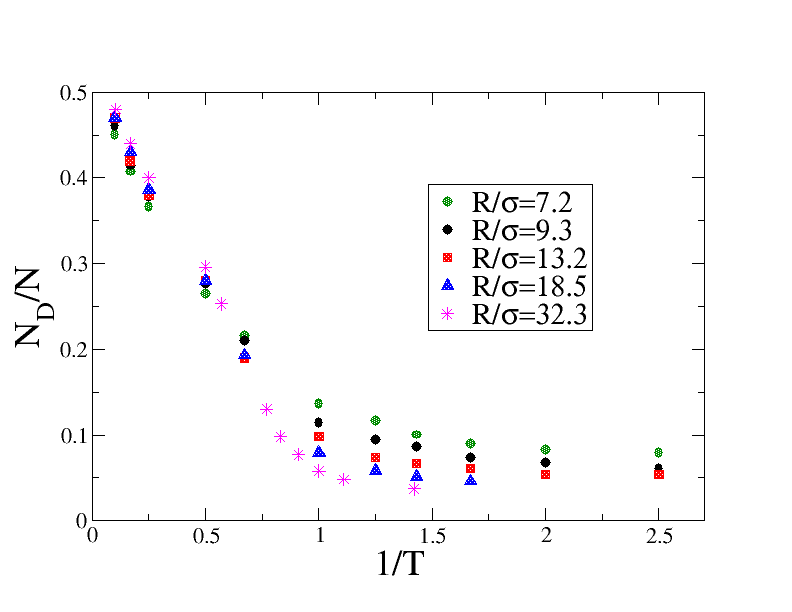}\hspace{2mm}
    \includegraphics[width=0.48\textwidth]{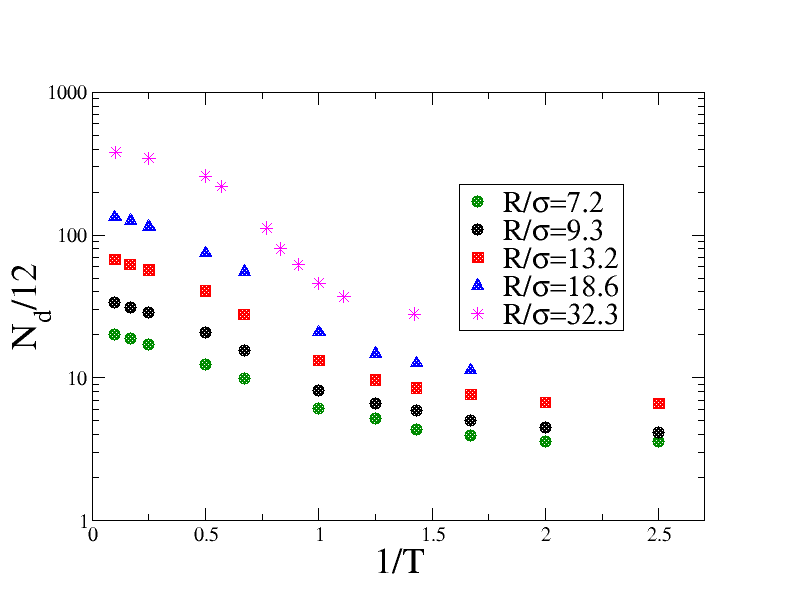}    
   \caption{\label{fig_ndefauts_fonc_T} Top: Total density of disclinations $N_D/N$ versus $1/T$ at $\tilde{\rho}=0.92$ and for  $5$ curvatures (see the inset). Bottom:  Ratio of the total number of dislocations divided by the irreducible number of curvature-induced disclinations, $N_d/12$, versus $1/T$ for the same curvatures. These two quantities decay with $1/T$ and go to a plateau value at low  $T$. }
\end{center}
\end{figure}

To make the observations more quantitative, we have computed the density of disclinations, $N_D/N$, (i.e. $+1$ and $-1$ disclinations), as well as  the number of dislocations per irreducible disclination (there are $12$ such disclinations on $S^2$), $N_d/12$, for five radii of curvature $R/\sigma=7.2$, $9.3$, $13.2$, $18.6$, and $32.2$. After an average over $1000$ configurations at equilibrium, the outcome is plotted versus the inverse of the temperature  in Fig.~\ref{fig_ndefauts_fonc_T}.

At high temperature, the density of disclinations $N_D/N$ has a value that is essentially independent of the curvature. Defects represent almost $50\%$ of the particles at the highest temperature and their density rapidly decreases with decreasing temperature. A dependence on $R/\sigma$ starts to be seen when $T$ approaches $T^*$, and the decrease of the disclination density is then stronger for smaller curvature ({\it i.e.}, larger $R/\sigma$). Around $T\approx 0.7$ and below, $N_D/N$ levels off and reaches a nonzero plateau value, whose value decreases with decreasing curvature. 

\begin{figure}[ht]
   \resizebox{8.5cm}{!}{\includegraphics{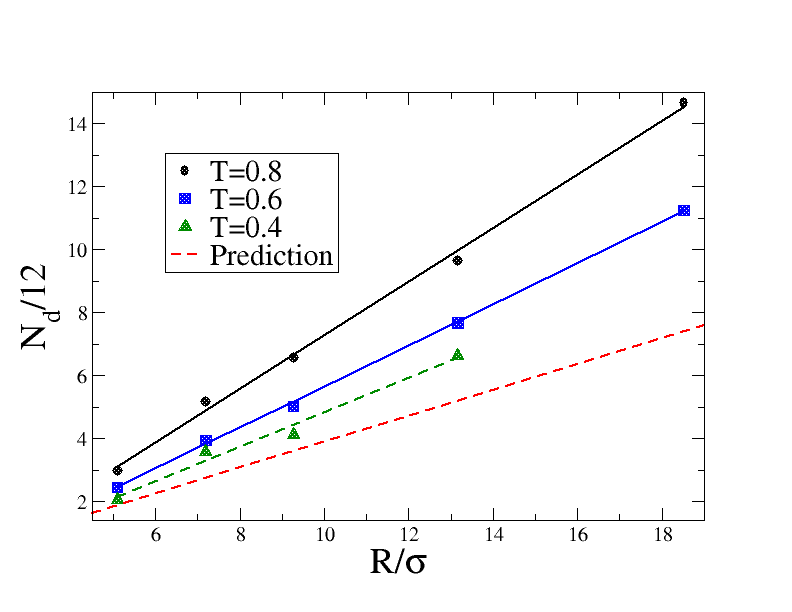}}
\caption{Low-temperature value of the number of dislocation per irreducible disclination, $N_d/12$, as a function of $R/\sigma$ (as obtained from Fig.~\ref{fig_ndefauts_fonc_T}. The dashed red line is the zero-temperature theoretical prediction from the continuum theory of defects \cite{Bowick2000}.}
\label{fig:dislocation_GBS}
\end{figure}

The low-$T$ saturation value is significantly larger than the minimum required by the topological constraint, {\it i.e.}, $12/N$. This is consistent with the existence of nontrivial structures of defects, the predicted grain boundary scars \cite{Bowick2000}, which persist at low $T$, possibly down to zero temperature. (This is true at least for the curvatures that we consider: For large curvatures, {\it i.e.}, small system sizes, one expects that the ground state is formed by simply $12$ positive $+1$ configurations at the vertices of a regular icosahedron \cite{Bowick2000,Bowick2009}.) These structures can be characterized by looking at the number of dislocations per irreducible disclination, $N_d/12$, which is plotted versus $1/T$ in the bottom panel of Fig.~\ref{fig_ndefauts_fonc_T}. The plot uses a semi-logarithmic scale because the variation of $N_d/12$ with  $R/\sigma$ at constant temperature is large. The number of dislocations per irreducible disclination decreases as one cools the system and reaches a value of the order of $10$ at low  temperature, the larger the curvature the smaller this value. We have plotted the low-$T$ values versus $R/\sigma$ in Fig.~\ref{fig:dislocation_GBS}. We find a roughly linear increase with $R/\sigma$ with a slope that decreases as temperature decreases and approaches the zero-temperature theoretical prediction given in Eq. (\ref{eq_GBSlength}). (It is hard to directly obtain the asymptotic value at zero temperature by using a nonlinear regression because the number of data points is small.)

\subsection{Further analysis of the defect configurations}

\begin{figure}[ht]
\begin{center}
   \resizebox{9cm}{!}
  {\includegraphics{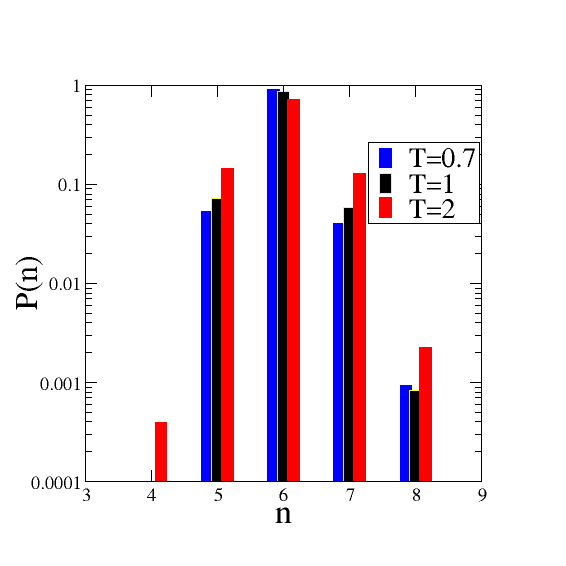}
    \includegraphics{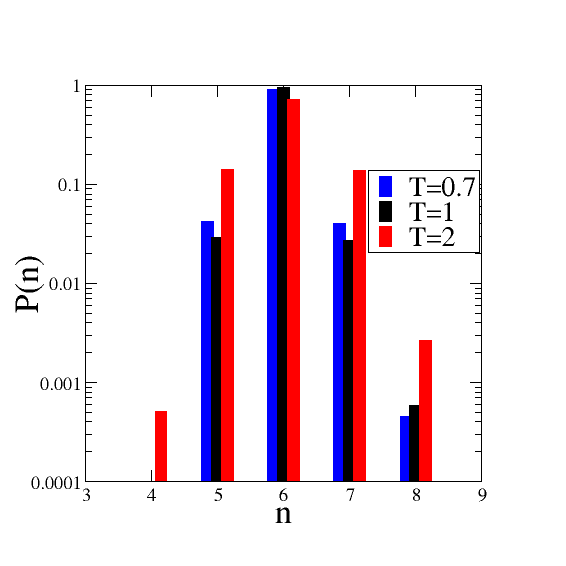}}
   \caption{\label{fig_histo_voisins} Linear-log plot of the distribution probability $P(n)$ of $n$-coordinated particles for  $\tilde{\rho}=0.92$, $T=2$, $1$, and $0.7$, in the case of two curvatures, $R/\sigma=9.3$ (left) and $R/\sigma=32.2$ (right). }
\end{center}
\end{figure}

In the above analysis, we have focused on the disclinations associated with $5$- and $7$-fold coordinated particles. We do see some configurations in which some $4$-coordinated and $8$-coordinated disclinations appear, but these defects are very rare because they are energetically unfavorable (no disclinations of higher charges are observed). This is confirmed by the histogram in Fig.~\ref{fig_histo_voisins}. For a temperature $T=1$ and below, there are no $4$-fold defects and the occurrence of $8$-fold ones is extremely small (the plot is a linear-log one).

We have seen by visual inspection in Fig.~\ref{fig:config_defauts} that the defects organize at low temperature in essentially $12$ grain boundary scars. To get a more detailed description of the defect organization, we have defined a localized defect structure as follows. We assume that two disclinations that are separated by a distance smaller than a cutoff $r_c=2\sigma$ belong to the same structure. This cutoff distance $r_c$ is somehow arbitrary, but is chosen such that  the mean number of disclination per structure matches the mean number of excess disclinations per irreducible disclination  $\big<n_{d}\big>/12$ at low temperature, which is consistent with the existence of $12$ grain boundary scars on average. We focus on the probability distribution of the number of disclinations, $n_{\text{c}}$, per independent localized structure. To compute this distribution, we first choose  randomly a dislocation in the system, which becomes a seed. For growing  the structure, we implement a depth-first search algorithm \cite{Even1979}. Starting from the seed, the two disclinations of the dislocation are labeled $1$ and $2$, and one searches all disclinations that are within a disk of radius $r_c$ of these labeled defects. The new disclinations are and also labeled. The search is iterated by using the new labeled disclinations as seeds. The structure is complete when no more disclinations can be added. A small fraction of defects then consists in isolated dislocations, but at low temperature, the algorithm provides $12$ large structures.

\begin{figure}[t]
\begin{center}
  \hspace{-0.2cm} \includegraphics[width=0.24\textwidth]{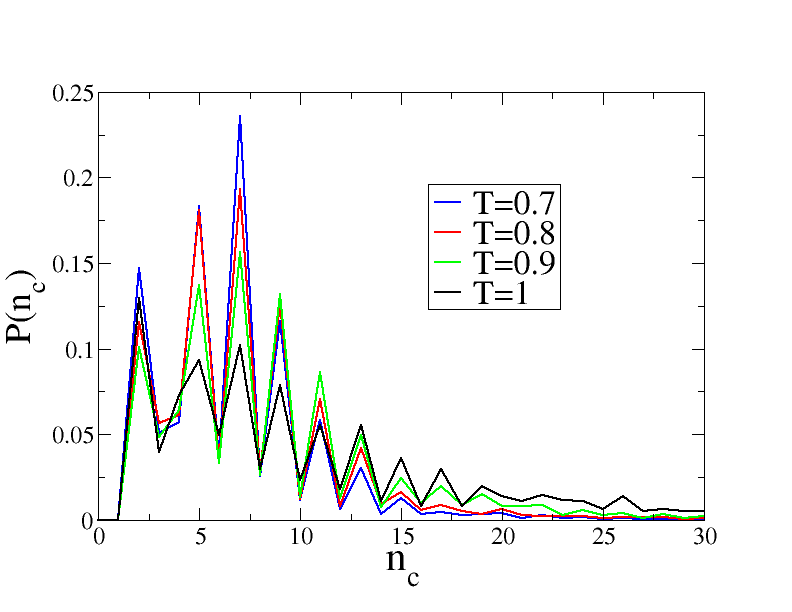}\hspace{-0.2cm}
    \includegraphics[width=0.24\textwidth]{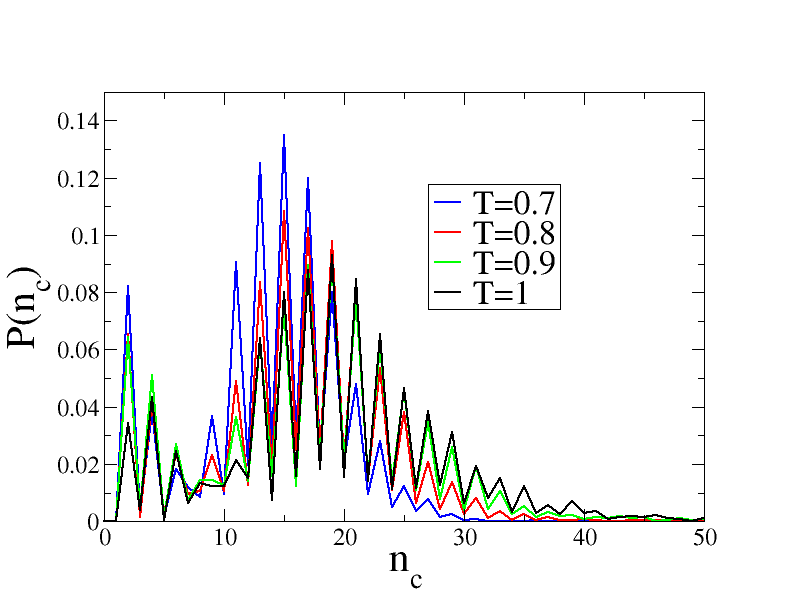}\\
   \hspace{-0.2cm} \includegraphics[width=0.23\textwidth]{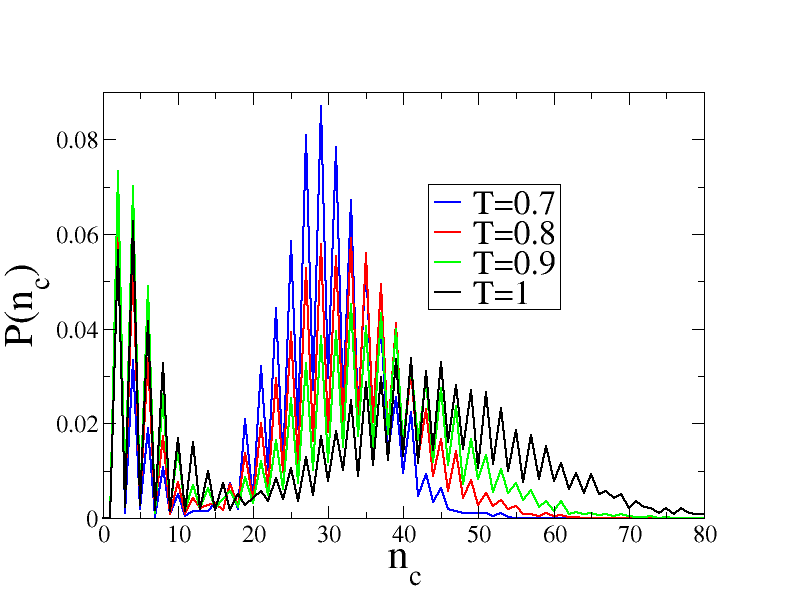}\hspace{-0.2cm}
    \includegraphics[width=0.23\textwidth]{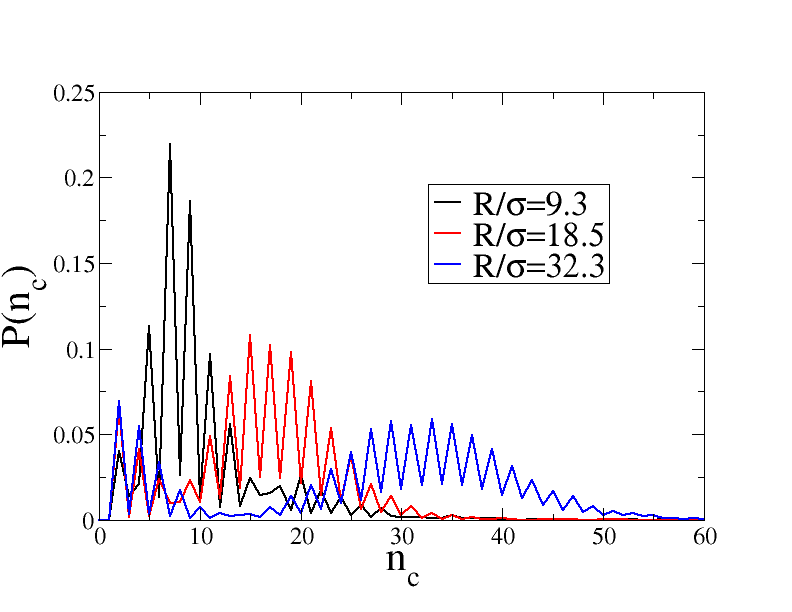}
   \caption{\label{fig_histo_scar} Probability distribution  $P(n_c)$ of the number of disclinations per localized defect structure: Evolution with temperature  for $R/\sigma=9.3$  (top left), $R/\sigma=18.6$ (top right), and $R/\sigma=32.2$ (bottom left). Bottom right: evolution with curvature for a fixed temperature, $T=0.8$.}
   \end{center}
\end{figure}

Fig.~\ref{fig_histo_scar} displays  $P(n_{\text{c}})$ for different temperatures, $T=1$, $0.9$, $0.8$, $0.7$,  and for $R/\sigma=9.3$ (top left), $R/\sigma=18.6$ (top right), $R/\sigma=32.2$ (bottom left). The bottom right panel shows $P(n_c)$ at fixed  $T=0.8$ for the three different radii of curvature. 

We note that $P(n_{\text{c}})$ exhibit saw-tooth profiles: $P(n_{\text{c}})$ is larger if $n_{\text{c}}$ is odd, {\it i.e.}, il there is an extra disclination in the defect structure as in the grain boundary scars formed by dislocations plus a $+1$ disclination, except for  $n_{\text{c}}=2$ which corresponds to an isolated dislocation. However, $P(n_{\text{c}})$ does not vanish if $n_{\text{c}}$ is even, which means that there may be ``closed'' structures built as a sequence of alternating disclinations $+1$ and $-1$. At low temperature, the distribution becomes peaked with a maximum that corresponds to the most probable value, and is larger the larger the ratio $R/\sigma$ as predicted by the continuum theory of defects and shown in Fig.~\ref{fig:dislocation_GBS}. Note that the typical and the mean values do not coincide especially at high $T$ because the distribution is asymmetric. Yet as $T$ decreases, the distribution significantly narrows.

\section{Results for the dynamics}
\begin{figure}[t!]
\begin{center}
 \resizebox{7.7cm}{!}{\includegraphics{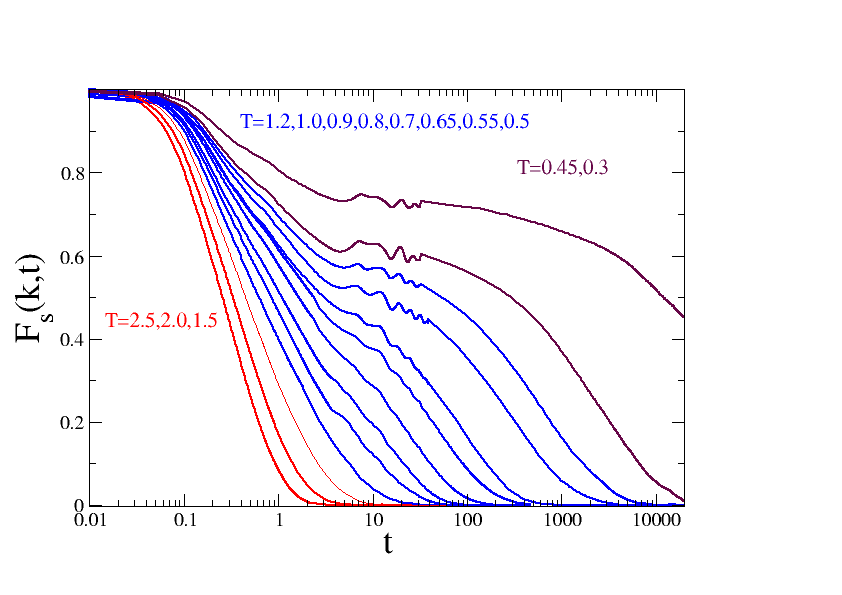}}\\
  \resizebox{7.7cm}{!}{\includegraphics{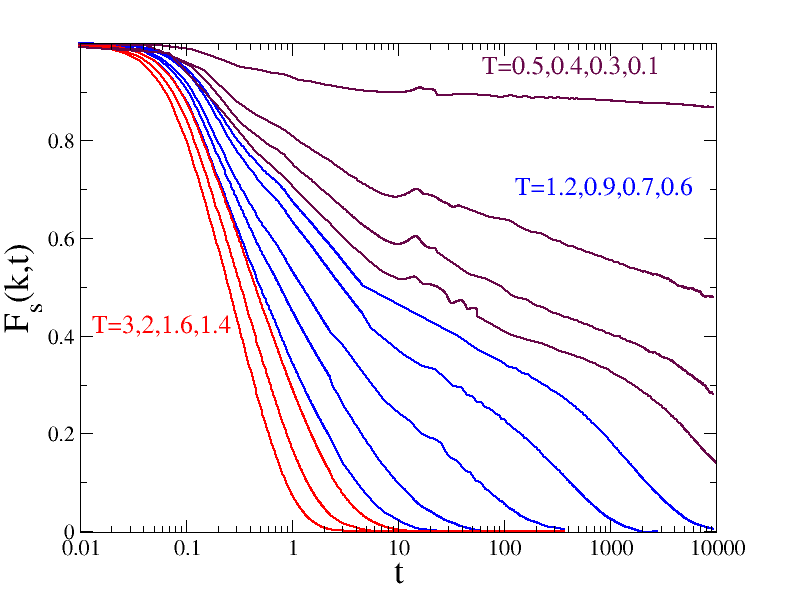}}\\
 \resizebox{7.7cm}{!}{\includegraphics{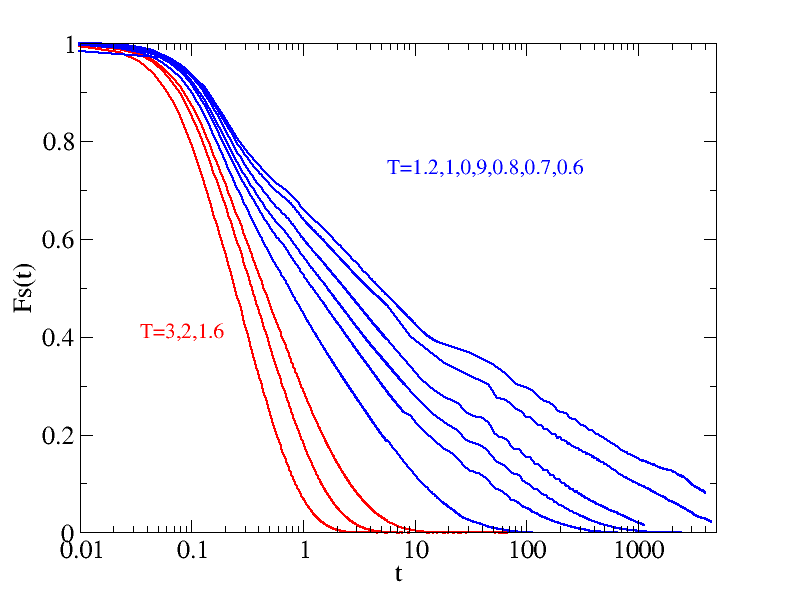}}
   \caption{\label{fig:Fsmultiple} Time dependence of the self intermediate scattering functions, $F_s(k,t)$, for several temperatures. The $3$ panels correspond to different curvatures:  $R/\sigma\approx 9.3$ (upper), $R/\sigma\approx 18.6$ (middle), $R/\sigma\approx  32.2$ (bottom). The red curves correspond to temperatures above the ordering transition $T^*$ in the Euclidean plane (``liquid'') and the blue curves are below but still equilibrated (``supercooled liquid''); the brown curves in the two upper  panels correspond to an out-of-equilibrium, aging, system (a ``glass'').}
\end{center}
\end{figure}
\subsection{Glassy dynamics: analysis of the intermediate scattering functions}

The main focus of our study is the dynamics of dense particle assemblies on a spherical substrate, and we consider the {\it equilibrium} dynamics, {\it i.e.}, the relaxation toward an equilibrated state.

 The structural relaxation of liquids and supercooled liquids is commonly characterized through the study of the self-intermediate scattering function $F_s(k,t)$ [see Eq. \ref{eq_Fs})] for $k$ corresponding to the maximum of the static structure factor $S(k)$ which is of the order of $2\pi$ over the typical distance between particles. The results are shown for several temperatures and several curvatures, $R/\sigma = 9.3,18.6$ and $ 32.2$ in Fig. \ref{fig:Fsmultiple}.  The curves show a strong slowdown of dynamics with decreasing temperature. The form of the function passes from a simple exponential at high temperature to a two-step decay at low temperature with the slowest part of the decay being describable by a stretched exponential, $\exp[-(t/\tau)^\beta]$ with $\beta <1$. 
 
 As anticipated, curvature thwarts crystallization and the system remains liquid down to temperatures much below the ordering transition in the Euclidean plane ($T^*\approx 1.3-1.4$). At some low temperature, we are not able to equilibrate the system in the time of the simulation (see the middle panel of Fig. \ref{fig:Fsmultiple} for $R/\sigma=18.6$ and $T=0.5,0.4,0.3$). The system can then be considered as a glass, {\it i.e.}, a slowly-aging frozen liquid or defective crystal. 

The dynamics of the particle assemblies on a spherical substrate therefore appear "glassy". We will explore later on several features of this glassiness. However, we note that, as also observed in other $2$-dimensional glass-forming atomic liquids and colloidal suspensions, the two step-decay does not seem to evolve at low enough temperature into a clearly established plateau marking a wide separation between the two relaxation regimes. This is to be contrasted with the behavior of glass-forming liquids in $3$ dimensions and will be further discussed in the following.

\begin{figure}[t]
\begin{center}
   \resizebox{12cm}{!}{\includegraphics{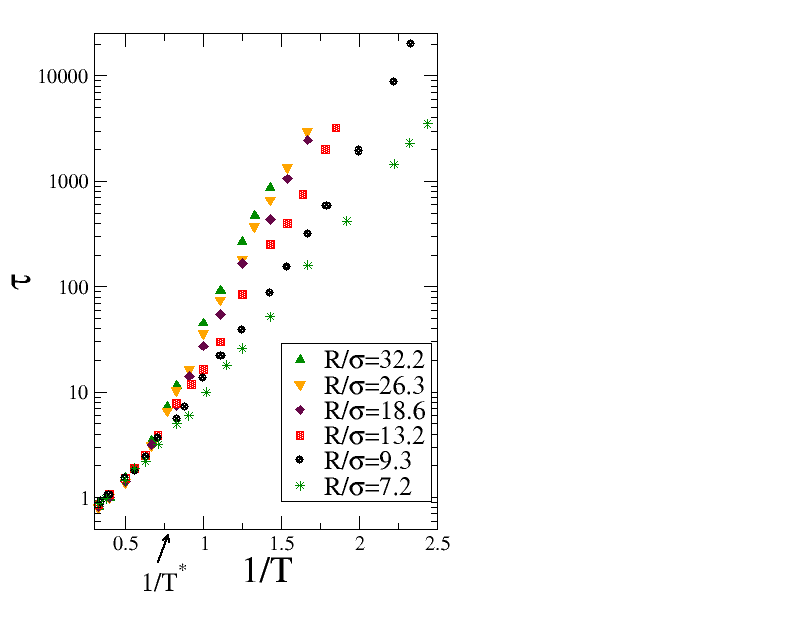}}
   \caption{\label{fig:tau_invtemp} Arrhenius plot of the temperature dependence of the relaxation time, $\log(\tau)$ versus $1/T$, for 6 different curvatures.  The arrow indicates the approximate location of the ordering transition(s) in the Euclidean plane.}
\end{center}
\end{figure}

From the self-intermediate scattering function $F_s(k,t)$ we extract a typical (alpha) relaxation time $\tau$. This can be performed through various more or less equivalent procedures, and we have chosen here to define $\tau$ as the time at which $F_s(k,t)=0.1$. (The different definitions of $\tau$ give different  values at a given temperature but lead to virtually the same temperature dependence.) The temperature dependence of the relaxation time is displayed in Fig. \ref{fig:tau_invtemp} in an Arrhenius plot, {\it i.e.}, as $\log\tau$ versus $1/T$, for 6 different values of the reduced radius of curvature: $R/\sigma \approx 7.2$, $9.3$, $13.2$, $18.6$, $26.3$, and $32.2$. At high temperature the relaxation time is essentially independent of curvature, as relaxation is only a local process for which all substrates therefore appear flat. Curvature starts to be felt around and below the ordering temperature in the Euclidean plane ($1/T^* \approx 0.72-0.77$), which is a signature of the avoided-transition phenomenon \cite{Kivelson1995}.

\begin{figure}[t]
\begin{center}
   \resizebox{8cm}{!}{\includegraphics{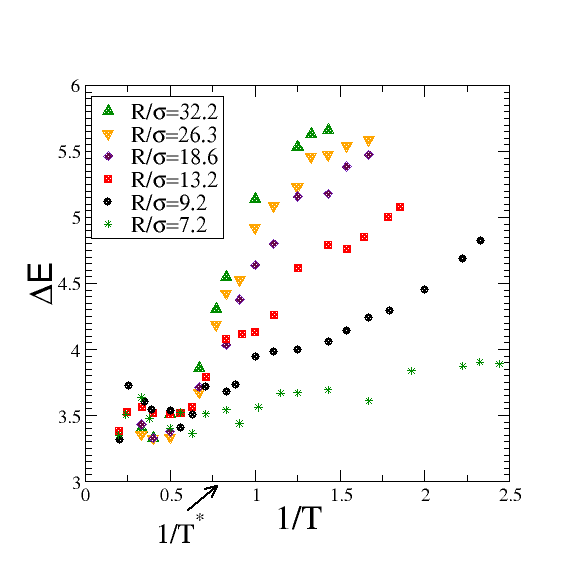}}
   \caption{\label{fig:activation_energy} Effective activation energy $\Delta E$ extracted from the Arrhenius plot of Fig. \ref{fig:tau_invtemp} as $\Delta E=T\ln[\tau(T)/\tau_0]$. Top: $\Delta E$ versus the inverse temperature $1/T$ for 6 different curvatures.  (The arrow indicates the approximate location of the ordering transition(s) in the Euclidean plane.) }
\end{center}
\end{figure}

\begin{figure}[t]
\begin{center}
    \resizebox{8.5cm}{!}{\includegraphics{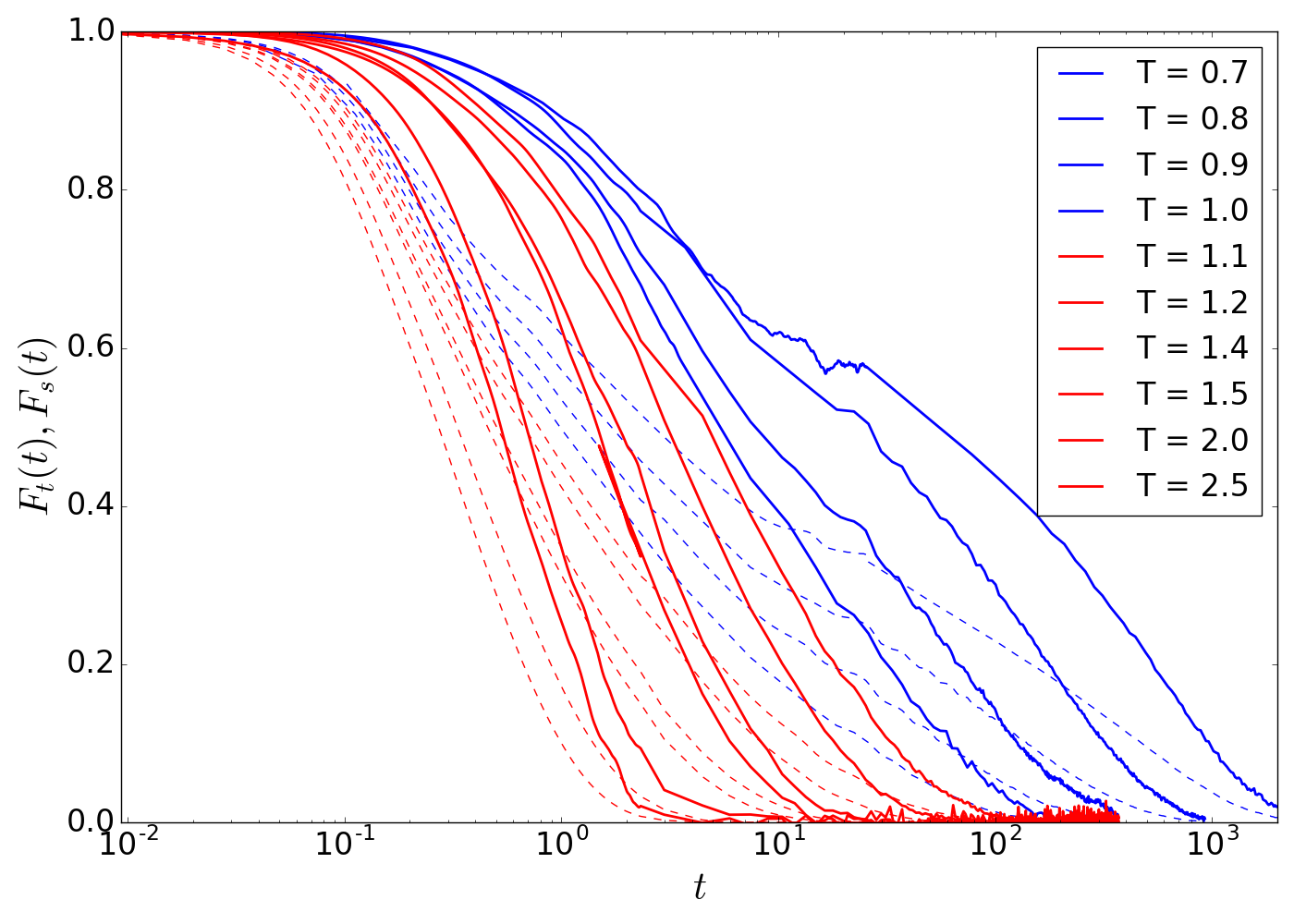}}
    \resizebox{8.5cm}{!}{\includegraphics{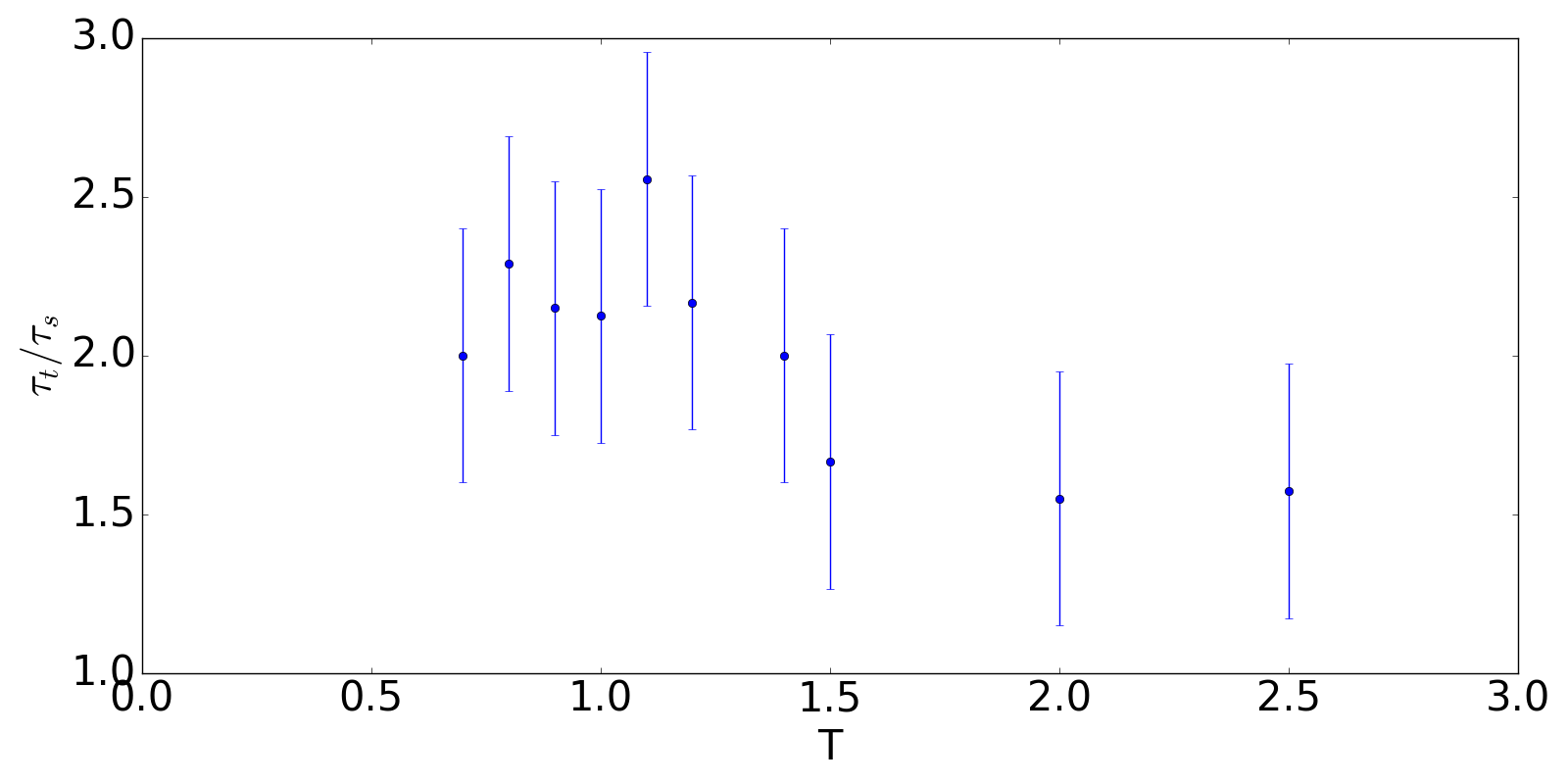}}
    \caption{\label{fig:FFs} (Top) Time dependence of the total and self intermediate scattering functions, $F(k,t)$ and $F_s(k,t)$, for different temperatures and for a curvature  $R/\sigma\approx 18.6$. (Bottom) Ratio of the relaxation times of $\tau_t/\tau_s$ of the total and self functions versus $T$.}
\end{center}
\end{figure}

The (alpha) relaxation time follows an Arrhenius dependence all the way to the lowest accessible temperatures for the largest curvature, $R/\sigma\approx 7.2$, but has a stronger increase as curvature decreases and rather displays a super-Arrhenius behavior, as observed in "fragile" glass-forming liquids. This stronger slowdown with decreasing curvature, {\it i.e.}, decreasing frustration, is also observed for the Lennard-Jones liquid on the negatively curved hyperbolic plane \cite{Sausset2008c,Sausset2010,Sausset2010a} and is predicted by the frustration-based theory of the glass transition \cite{Tarjus2005}. One however finds that the super-Arrhenius increase appears to somehow saturate for the smallest curvatures with a possible crossover to an Arrhenius dependence at the lowest temperatures. Whether this phenomenon is a consequence of the finite size of the system (which is of course a real physical phenomenon on the sphere) or is due to the difficulty to properly equilibrate the liquid at the lowest temperatures (see also below) is still unclear.

In Fig. \ref{fig:activation_energy} (top panel) we have plotted the temperature dependence of the effective activation energy, obtained from $\Delta E(T)=T\ln[\tau(T)/\tau_0 ]$ where $\tau_0$ is a high-temperature relaxation time. The figure illustrates the curvature-independent Arrhenius dependence observed at high-$T$ and the significant deviation from this dependence as $T$ decreases, except for the largest curvature $R/\sigma\approx 7.2$ for which the effective activation energy is essentially constant over the whole temperature range. At low temperature there is a trend towards an increase of the activation energy with decreasing curvature. Howeber the error bars are too large to significantly test the linear dependence on $R/\sigma$ predicted by the dislocation glide description, Eq. (\ref{eq_delta_E_dislo}).

We have also considered the total intermediate scattering function, $F(k,t)$ [see Eq. (\ref{eq_Fcoll})], with as before $k$ corresponding to the maximum of the static structure factor. This function corresponds to the pair correlations of the fluctuations of the total particle density at the  wavevector $k$. For $k$ probing typical inter-particle distance in liquids and supercooled liquids, the total and the self intermediate scattering functions usually show a very similar behavior. One may wonder however in the present case where large, albeit mobile and transient, hexagonal domains form at low temperature whether slower relaxation modes than the structural one seen in  $F_s(k,t)$ could affect $F(k,t)$. As illustrated in Fig. \ref{fig:FFs} for $R/\sigma\approx 18.6$ and several temperatures, the decay of $F(k,t)$ is indeed significantly slower than that of $F_s(k,t)$ but by less than an order of magnitude: For all temperatures,  the ratio of relaxation times of the total and self functions remains close to a factor $2$ (see the bottom panel of Fig. \ref{fig:FFs}). This is enough to make the full equilibration of the system harder but does not seem to indicate a qualitatively new relaxation channel.


%

\subsection{Particle trajectories: evidence for growing dynamical intermittency and dynamical heterogeneity}

\begin{figure}[ht]
\begin{center}
 \resizebox{8.9cm}{!}{ 
     \hspace{-0.1cm}\includegraphics{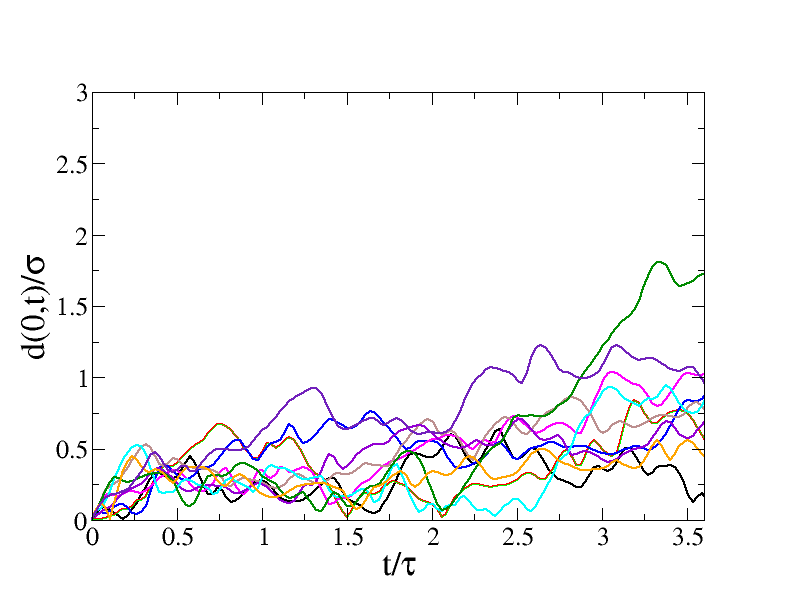}\hspace{-0.1cm}
     \includegraphics{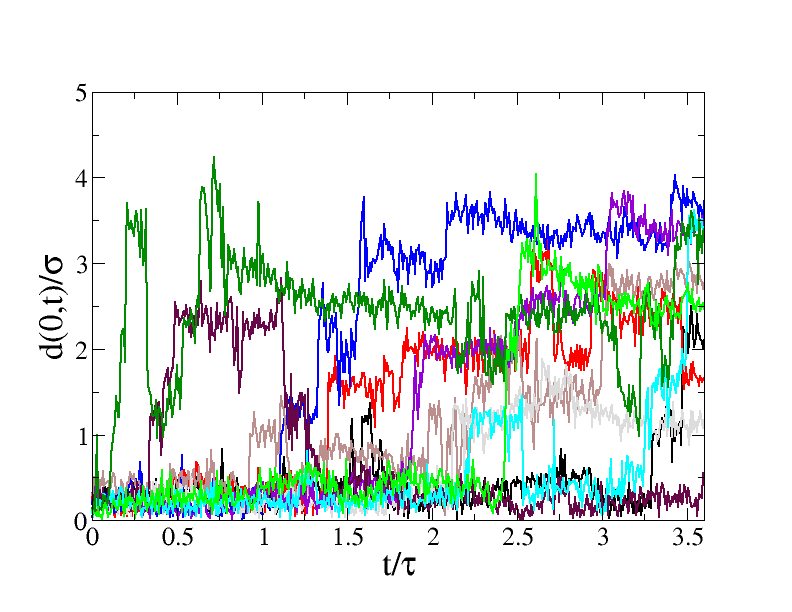}
     }
    \caption{\label{fig_trajectories_fonc_T}  Evolution of the geodesic distance  $d(0,t)/\sigma$ traveled by several randomly chosen  particles as a function of the  reduced time $t/\tau$ for  $R/\sigma \approx 9.3$. Two temperatures are shown: left, $T=2$; right,  $T=0.5$ (recall that the avoided ordering transition $T^*\approx 1.3-1.4$).}
   \end{center}
\end{figure}

\begin{figure}[h!]
\begin{center}
\includegraphics[width=8.8cm]{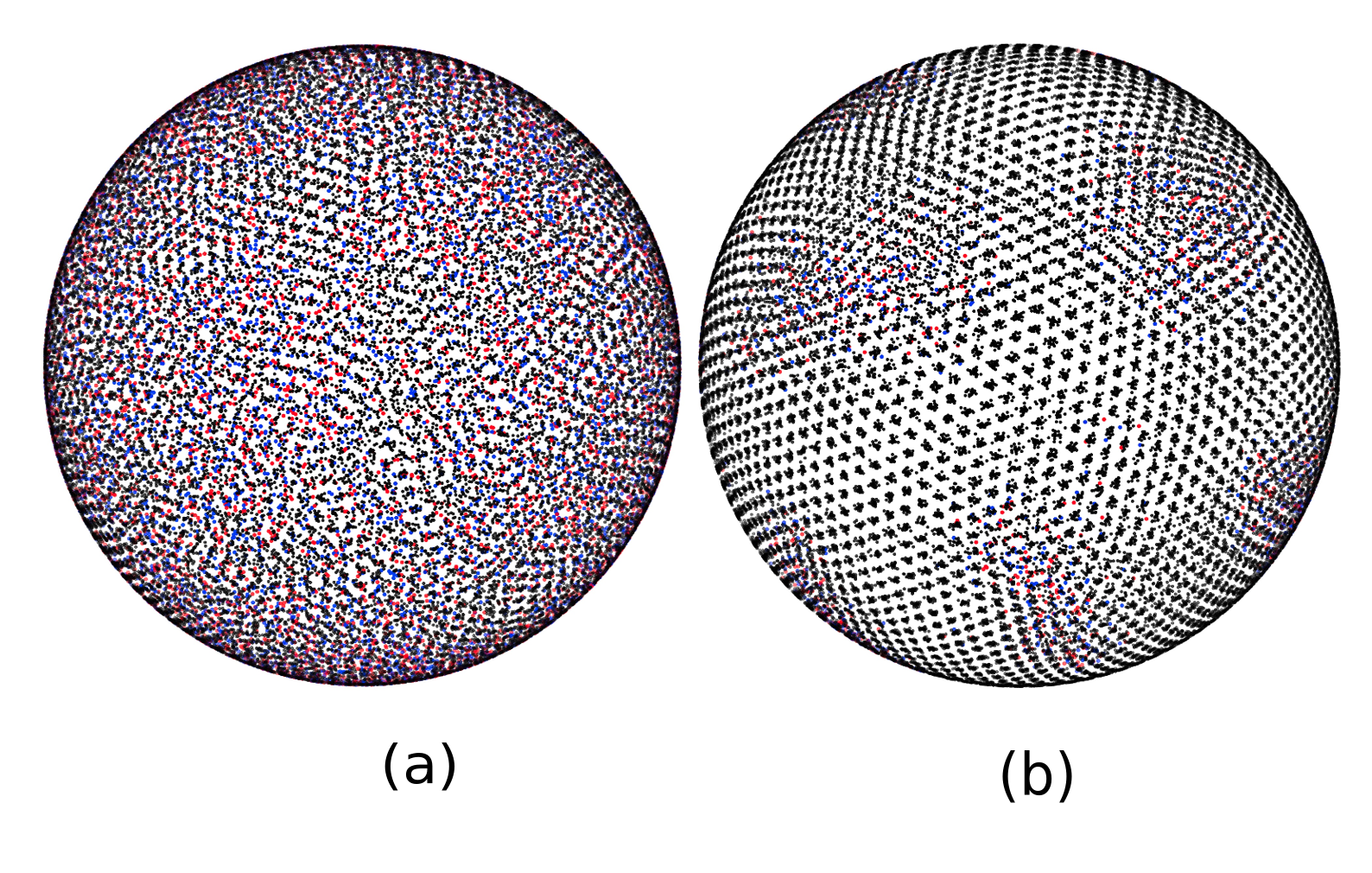}
\caption{\label{fig_config_superposees} Particle trajectories followed for a time interval during which the particles have traveled over an average geodesic distance equal to $\sigma$, for $R/\sigma \approx 18.6$, and two temperatures (a) $T=2$, (b) $T=0.7$ (recall that the avoided ordering transition $T^*\approx 1.3-1.4$).  At $T=2$, all particles move and the system remains homogeneous. At $T=0.7$, particles move very little in the large hexatic/hexagonal regions (black) whereas significant particle motion is present in regions with defect structures (red).}
\end{center}
\end{figure}

We have monitored the displacement of the particles, as well as the number of their nearest neighbors, with time.  In Fig. \ref{fig_trajectories_fonc_T} we illustrate the geodesic distance traveled by several randomly chosen  particles as a function of the rescaled time $t/\tau$ for a dimensionless radius of curvature $R/\sigma\approx9.3$ and two temperatures, one significantly higher than the avoided transition temperature $T^*$ and one significantly lower. We observe the common signature of a glassy slowdown of dynamics, {\it i.e.}, the passage from a rather homogeneous behavior at high temperature to an intermittent and strongly particle-dependent behavior at low temperature. This can be visualized at the level of the whole system by overlaying snapshots of particle configurations corresponding to an elapsed time during which the mean geodesic distance between the initial and final positions of the particles is equal, say, to $\sigma$: see Fig. \ref{fig_config_superposees}. At high temperature ($T=2$) all particles move by a comparable amount and the dynamics is spatially homogeneous. At low temperature ($T=0.7$) on the other hand, most  particles hardly move or just rattle in the cage formed by their neighbors and mobility is concentrated in rare localized  regions,  which illustrates the spatial heterogeneity of the dynamics. This heterogeneity is highly correlated with the presence of topological defects. One finds by inspection that at low temperature the  regions  of  high  mobility  coincide  with the grain boundary scars and their vicinity whereas the regions of low mobility coincide with the hexagonal domains. These results are similar to what has been found for the same liquid on the hyperbolic plane $H^2$. For this reason we have not repeated the detailed characterization of the growing dynamical heterogeneity performed on $H^2$ by computing $4$-point dynamical quantities \cite{Sausset2010}. A comparable observation has also been recently made in the experimental study of a monolayer of colloidal particles on spherical droplets \cite{Guerra2018a}.

\begin{figure}[t]
\begin{center}
 \resizebox{8.5cm}{!}{ 
    \includegraphics{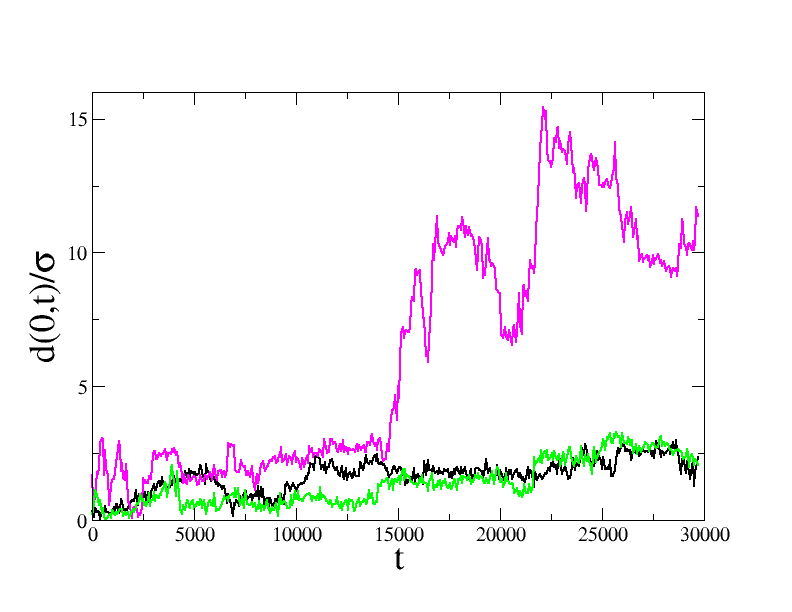}
     }\\
    \resizebox{8.5cm}{!}{ 
    \includegraphics{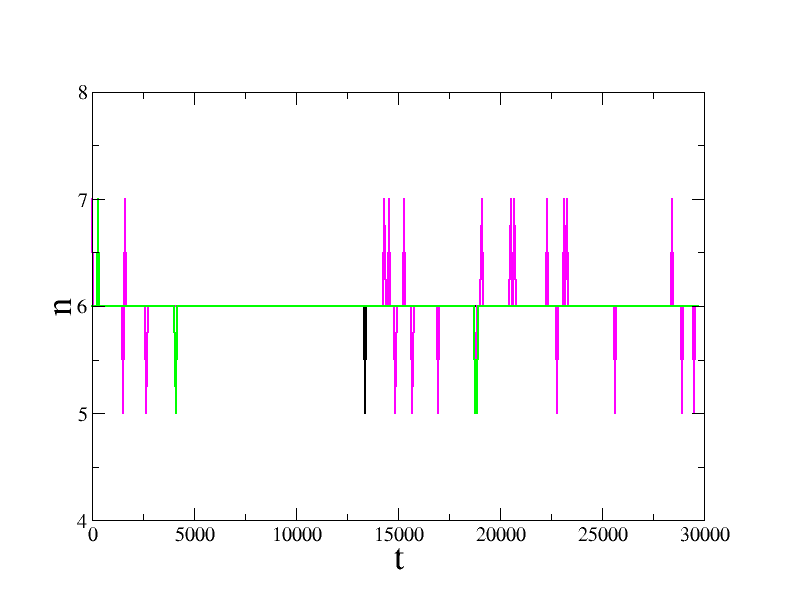} }
   \caption{\label{fig_nvoisins_fonc_T} (Top) Time evolution of the geodesic distance  $d(0,t)/\sigma$ traveled by three  particles for 
   $R/\sigma \approx 18.6$ and $T=0.7$. The elapsed time corresponds to  $60 \tau$ (much larger than in Fig. \ref{fig_trajectories_fonc_T}). (Bottom) Time evolution of the coordination number $n$ of the same $3$ particles. }
   \end{center}
\end{figure}

We have also followed the particle trajectories over very long times, much larger than the relaxation time $\tau$, to verify that the particles in the $6$-fold (hexagonal or hexatic) domains are nonetheless moving over significant distances and investigate in more detail the connection between mobility and local environment (mainly, coordination number). 
In Fig. \ref{fig_nvoisins_fonc_T} we illustrate the results for 3 particles followed over an elapsed time of  $60 \tau$ for the same system as in Figs. \ref{fig_trajectories_fonc_T} and  \ref{fig_config_superposees}, {\it i.e.}, $R/\sigma\approx 18.6$, and for a temperature $T=0.7$ below $T^*$. One of the three particles moves over a much larger distance than the two others: it moves over distances slightly less but of the order of $R$ while the other two move over a few particle diameters $\sigma$. When looking at the right panel of Fig. \ref{fig_nvoisins_fonc_T}, one can see that the latter two particles stay virtually all the time $6$-fold coordinated; on the other hand, the  other particle spends a significant fraction of its trajectory being a disclination with either $5$ or $7$ nearest neighbors, especially when it is more mobile (for $t\gtrsim 1500$). This connects again particle mobility with topological defects. One notes however that the most mobile particle does not remain a defect at all time. It switches rapidly between being a disclination and being $6$-fold coordinated and it spends even more time in the latter situation. This short lifetime of the identification of a disclination with a specific particle makes it hard to monitor the dynamics of individual defects. (For this matter, this is different from grain boundaries in true polycrystalline materials, see {\it e.g.}, Ref. [\onlinecite{Zhang2009}].) A particle indeed changes its coordination number under the displacements of its nearest neighbors, displacements that can be small and nonetheless efficient in locally altering the Voronoi construction. 

\subsection{Mean weighted displacements}

To try to characterize the displacement of the particles as a function of their local topology, {\it i.e.} their  coordination number, one must therefore account for the rapid changes of coordination number, especially when the latter differ from the mean and typical (see Fig. \ref{fig_histo_voisins}) value, $6$. To do so, we have defined a mean weighted displacement for $n$-fold coordinated particles, with $n=5$ (positive disclination), $6$, and $7$ (negative disclination), 
between two times $0$ and $t$ as
\begin{equation}
d_n(0,t)=\frac{1}{N_n}\bigg<\sum_{i=1}^N d_i(0,t)\Big [\frac{1}{t}\sum_{t''=0}^t\delta_{n_i(t"),n}\Big ]\bigg>,
\end{equation}
where $N_n$ denotes the mean number of $n$-fold coordinated particles during the simulation, $d_i(0,t)$ the geodesic distance between the positions of the particle $i$ at times  $t=0$ and  $t$,  $n_i(t)$ is the coordination number of particle $i$ at time $t$, and $\delta_{n',n}$ is a Kronecker symbol. The weighting factor represents the fraction of the elapsed time during which the chosen particle is $n$-fold coordinated.

In the case of the $6$-fold coordinated particles we have also studied a more restrictive mean weighted displacement, $d_{66}(0,t)$, in which only $6$-coordinated particles that are surrounded by $6$ particles of coordination number $6$ are taken into account. This quantity better characterizes the mobility of particles that are in the bulk of hexagonal (or hexatic) domains.

\begin{figure}[ht!]
\begin{center}
 \hspace{-0.2cm}\includegraphics[width=0.24\textwidth]{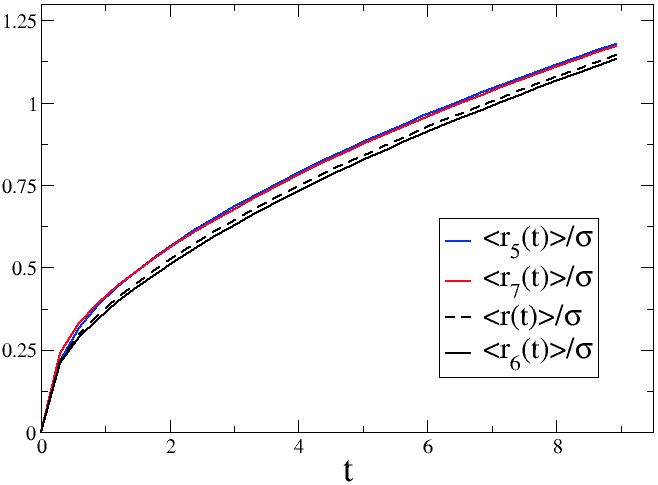}
 \includegraphics[width=0.24\textwidth]{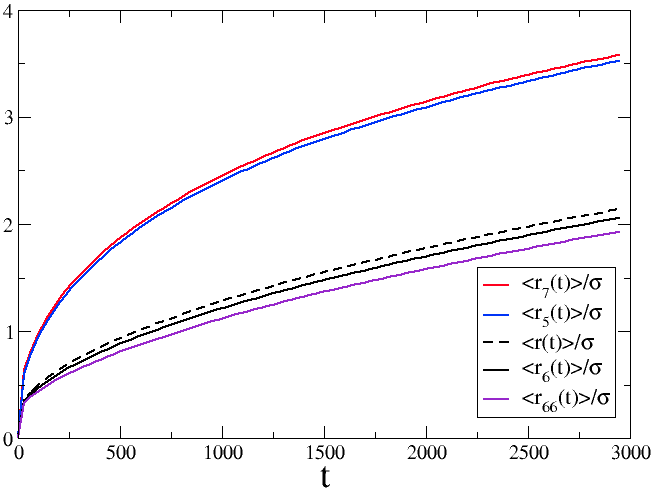}\\
 \hspace{-0.2cm}\includegraphics[width=0.24\textwidth]{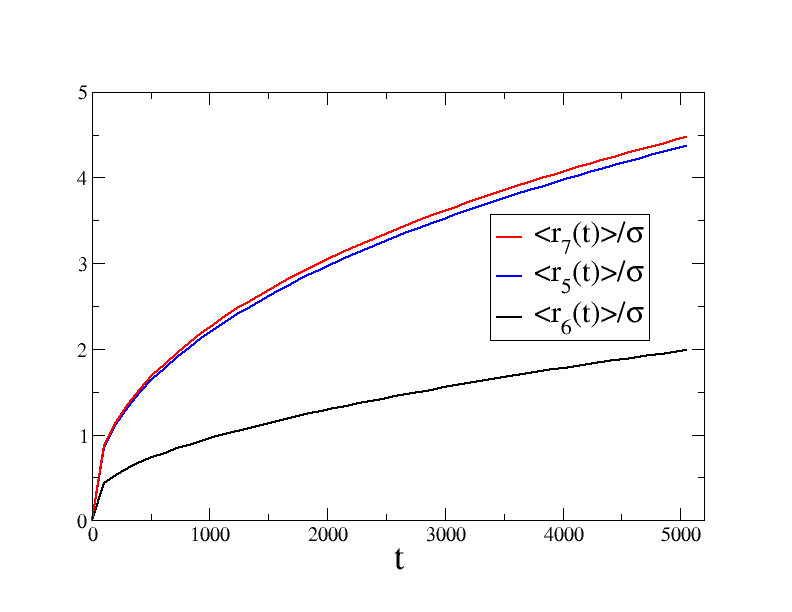}
 \includegraphics[width=0.24\textwidth]{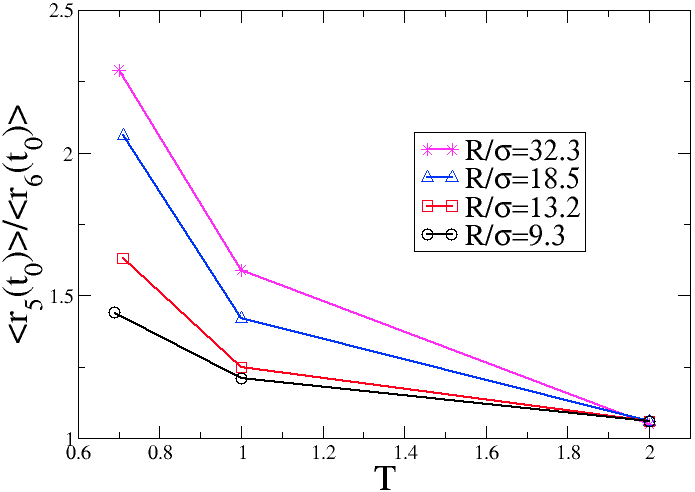}
\caption{\label{fig_r657} Mean displacement  $\big<d(0,t)\big>$ (dashed line) and mean weighted displacements for an elapsed time $t$ of about  $6\tau$ ()except for the bottom right panel. Top left:  $\big<d_6(0,t)\big>$ (full line), $\big<d_5(0,t)\big>$ (blue), $\big<d_7(0,t)\big>$ (red) for  for $R/\sigma\approx18.6$ and $T=2$. Top right: Same quantities and  $\big<d_{66}(0,t)\big>$ (violet) for   for $R/\sigma\approx18.6$ and $T=0.7$.  Bottom left: $\big<d_5(0,t)\big>$, $\big<d_7(0,t)\big>$, and $\big<d_6(0,t)\big>$ for $R/\sigma\approx 32.2$ and $T=0.7$.  Bottom right: Ratio $\big<d_5(0;t)\big>/\big<d_6(0,t)\big>$ versus $T$ for several values of $R/\sigma$ and an elapsed time of $3\tau$.}
\end{center}
\end{figure}

We show the mean weighted displacements, $d_{5}(0,t)$, $d_{7}(0,t)$, $d_{6}(0,t)$, $d_{66}(0,t)$, together with the total mean displacement, $d(0,t)$,  for an elapsed time $t$ up to $6 \tau$ and for different values of the temperature and the curvature in Fig.~\ref{fig_r657}.  
At high temperature (top left panel), there is virtually no difference between all the curves but at low temperature below $T^*$ (top right and bottom panels) there is a clear distinction between the mean weighted displacements of the defects and the others: At $T=0.7$ the disclinations have a higher mobility than the $6$-fold coordinated particles, their displacement being roughly twice bigger than that of the $6$-coordinated particles after some short transient time of O($1$). (This ratio increases with decreasing curvature, but the effect is rather moderate, as shown by the bottom right panel.) This higher mobility is expected from the previous observations. As seen from the top right panel, the displacements $d_{6}(0,t)$ and $d_{66}(0,t)$ are very similar, which indicates that the distinction between edges and bulk for $6$-fold ordered regions has no significant signature in the dynamics. Note also that at low $T$ the number of defects is small so that the average particle displacement is essentially given by that of the $6$-fold coordinated particles. An additional interesting observation is that the mobility of the $5$-fold and $7$-fold disclinations is essentially the same, even at low temperature. This is compatible with the fact that except for 12 irreducible $5$-fold disclinations all the other disclinations pair to form dislocations and each of these $5-7$ dipoles then move as a single defect. However, the dynamics seems to involve more than just dislocation glide in an otherwise frozen environment.

\subsection{Discussion: transient caging and the effect of long wavelength fluctuations}

\begin{figure}[t]
\begin{center}
    \resizebox{8cm}{!}{\includegraphics{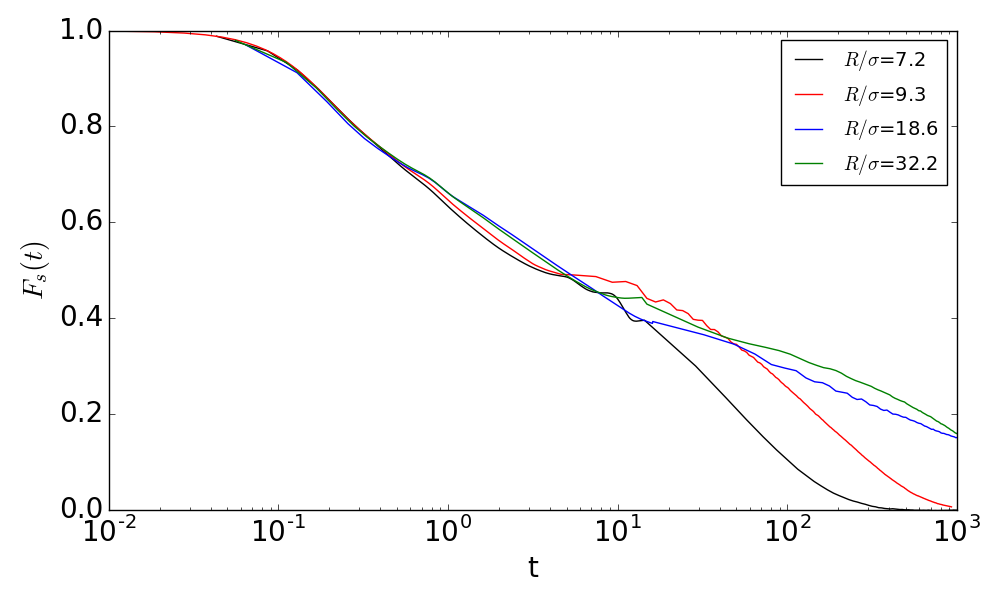}}
\caption{  Time dependence of the self intermediate scattering functions, $F_s(k,t)$, for $T=0.6$ and 4 different curvatures:  $R/\sigma\approx 7.2$ ($N=600$), $R/\sigma\approx 9.3$ ($N=1000$), $R/\sigma\approx 18.6$ ($N=4000$), and $R/\sigma\approx 32.2$ ($N=12000$). We focus on the intermediate time region.}
\label{fig:plateau}
\end{center}
\end{figure}

We have seen that the intermediate scattering functions at low temperature do not show a well-developed plateau separating the short- and the long-time relaxation regimes. The plateau which is found in $3$-dimensional glass-forming liquids is associated with a transient caging phenomenon in which particles are trapped for long times in the ``cage'' formed by their neighbors. Further relaxation of the system then requires a reorganization of the local environment. Strong evidence has been provided that $2$-dimensional glass-forming systems, either liquids studied by computer simulation \cite{Flenner2015,Shiba2016} or colloidal suspensions studied experimentally \cite{Vivek2017,Illing2017}, do not display a well-established plateau and that the behavior at intermediate times of their self intermediate scattering function shows a strong dependence on system size: a plateau is observed for small system sizes and gradually disappears as one increases the system size. Our present results on the $2$-sphere $S^2$ are not as clear-cut. As one lowers the temperature, the tendency to develop a plateau appears stronger for $R/\sigma\approx 9.3$ ($N=1000$) than for the larger systems (see Figs. \ref{fig:Fsmultiple} and \ref{fig:plateau}), but the smaller system $R/\sigma\approx 7.2$ ($N=600$) barely has a shoulder in the intermediate-time regime. Furthermore, the (alpha) relaxation time increases as the curvature decreases and the system size increases, so that Fig. \ref{fig:plateau} does not look like the $2$-$d$ flat-space system-size dependence displayed in Fig. 3 of Ref. [\onlinecite{Flenner2015}]. This illustrates that on the $2$-sphere the finite-size effect is mixed with that of curvature.

The progressive disappearance of a plateau with increasing system size in $2$-dimensional glass-formers has been taken as an indication that there are long wavelength modes allowing the particles to move along with their neighbors with no change of the local environment (and no ``cage breaking'') \cite{Vivek2017,Illing2017}. These long wavelength (density) modes have been termed Mermin-Wagner modes, as they destroy long-range periodic order in $2$ dimensions; they have been argued to play a role in glasses and amorphous solids at large length scale and on the dynamics in glass-forming liquids \cite{Vivek2017,Illing2017,Tarjus2017}. They may indeed provide an additional channel to relaxation on top of the reorganization of the local environment that is usually associated with the ``structural'' relaxation and glass formation.

An interesting question that would be worth further investigation is therefore to understand in more detail the role of the Mermin-Wagner-like density fluctuations in particle assemblies on a uniformly curved substrate. In frustrated systems these long wavelength fluctuations associated with the long-range extension of the local order are indeed limited by some intrinsic frustration length (here, the radius of curvature), and at low temperature their effect is intertwined with that of the topological defects \footnote{ One can compare liquids in uniformly curved $2$-dimensional manifolds with the $2$-dimensional uniformly frustrated XY model \cite{Esterlis2017}: In the latter model too, there is an avoided ordering (ferromagnetic) transition due to frustration and a relaxation slowdown when decreasing the temperature below the characteristic temperature of this avoided transition. The temperature dependence of the relaxation time is however quite different, with no super-Arrhenius behavior and instead a regime with sub-Arrhenius behavior. It has been argued in Ref. [\onlinecite{Esterlis2017}] that the difference with liquids in curved space comes from the nature of the degrees of freedom. In the XY model the relevant fluctuations are spin waves (corresponding to Mermin-Wagner-like soft modes) and localized topological defects in the form of vortices. Liquids in $2$ dimensions have long wavelength density modes but two types of topological defects, disclinations and dislocations that correspond to two types of order, bond-orientational and positional. While one may associate spin waves with density fluctuations and vortices with disclinations, dislocations have no real analog in the frustrated XY mode. This may explain the difference found in the dynamical behaviors of the two types of uniformly frustrated systems.}. Contrasting the behavior on the hyperbolic plane, where radius of curvature and system size are independent, and on the sphere, where radius of curvature and system size are linked, would be valuable. In addition, to probe the genuine glassy dynamics and possibly observe the same phenomenology as in $3$-dimensional glass-forming liquids, one could investigate observables that are somehow insensitive to the long wavelength fluctuations and more directly detect the dynamics of the rearrangements of the local environments. This seems to be the case of the time-dependent generalization of the bond-orientational correlation function, $G_6(r,t)$,  and of correlation functions that only involve the breaking of the local cage formed by the nearest neighbors (see Refs. [\onlinecite{Vivek2017,Illing2017,Flenner2015,Shiba2016,Turci2017}]).
This is however out of the scope of the present work and is left for a future study.

\section{Conclusion}

We have studied the behavior of a dense one-component system of particles constrained on a spherical substrate, a model that is relevant to describe  particle-stabilized (Pickering) emulsions. Our main focus has been the dynamics and the relaxation to equilibrium of the system as one varies the control parameters: temperature and curvature. We have shown that the dynamics is glassy, in that it displays many of the characteristic features of glass-forming liquids: a strong slowdown of the relaxation as temperature decreases and the emergence of a spatially heterogeneous dynamics. These features are more developed as the curvature of the substrate decreases. 

The prevalent $6$-fold local order is frustrated by the curvature and the configurations at low temperature appear as hexagonally/hexatically ordered regions interrupted by topological defects that form grain boundary scars. These configurations have been described as ``spherical crystals''\cite{Bausch2003,Einert2005,Lipowsky2005,Bowick2009} or rather as some form of defect-ordered phase where the $12$ localized defect regions permanently arrange at the vertices of an icosahedron \cite{Guerra2018a}. They have been proposed as potential candidates for technological applications such as defect functionalization. What we show here by varying the temperature (but the same phenomenology would be observed by decreasing the density) is that, when they appear solid-like on the observation time scale, these configurations can also be considered as ``glasses''. One therefore expects that they display some nontrivial out-of-equilibrium dynamics such as aging, which can be of relevance in practical applications.


%
\end{document}